\documentclass[12pt]{article}

\usepackage{amsmath,amsfonts,amssymb,amsthm}
\usepackage{color, graphicx}
\usepackage{epsfig,fullpage}
\usepackage{url}
\usepackage{lineno}
\usepackage{subfigure}
\usepackage{mathtools,MnSymbol}
\usepackage{soul}
\usepackage[blocks]{authblk}
\usepackage{multirow}
\usepackage{natbib}

\newcommand{\BEAS}{\begin{eqnarray*}} 
\newcommand{\EEAS}{\end{eqnarray*}} 
\newcommand{\BEA}{\begin{eqnarray}} 
\newcommand{\EEA}{\end{eqnarray}}
\newcommand{\BIT}{\begin{itemize}} 
\newcommand{\EIT}{\end{itemize}}
\newcommand{\BNUM}{\begin{enumerate}} 
\newcommand{\ENUM}{\end{enumerate}}

\newcommand{\ie}{{\it i.e.}$~$}

\newcommand{\bR}{{\bf R}}

\newcommand{\bS}{{\bf S}}

\newtheorem{theorem}{Theorem}

\newtheorem{proposition}{Proposition}
\newtheorem{algorithm}{Algorithm}

\begin{document}

\title{Testing Independence of Bivariate Censored Data using Random Walk on Restricted Permutation Graph}

\renewcommand\Affilfont{\itshape\small}
\author{Seonghun Cho, Donghyeon Yu, Johan Lim
\thanks{S. Cho is at Department of Statistics, Iowa State University, IA, USA. J. Lim is at Department of Statistics, Seoul National University, Seoul, Korea. D. Yu is 
at Department of Statistics, Inha University, Incheon, Korea. All correspondence are to J. Lim, \texttt{E-mail:johanlim@snu.ac.kr}.}}

\date{} 
\maketitle 

\begin{abstract}

\noindent 
In this paper, we propose a procedure to test the independence of bivariate censored data, which is
generic and applicable to any censoring types in the literature. 
To test the hypothesis, we consider a rank-based statistic,
Kendall's tau statistic. The censored data defines a restricted permutation space
of all possible ranks of the observations. We propose the statistic, the average of Kendall's tau over the ranks in the restricted permutation space. To evaluate the statistic and its reference distribution, we develop
a Markov chain Monte Carlo (MCMC) procedure to obtain uniform samples on the restricted permutation space and numerically approximate the null distribution of the averaged Kendall's tau. We apply the procedure to three real data examples with different censoring types, 
and compare the results with those by existing methods. We conclude the paper with some additional discussions not given in the main body of the paper.

\end{abstract}

\baselineskip 24pt 

\section{Introduction}

Let $\left( \mathbf{X}^{\ast}, \mathbf{Y}^{\ast} \right) = \left\{ \left( X_{i}^{\ast}, Y_{i}^{\ast} \right) : i = 1,\ldots,n \right\}$ be complete bivariate variables
of interest. There are many circumstances where the complete data are not available, which include
right or interval censoring, missing observations, micro-aggregation or local suppression for data privacy, and etc. In this paper, we particularly
focus on incompleteness by censoring. Let $\left( \mathbf{X}, \mathbf{Y} \right) = \left\{ \left( X_{i}, Y_{i} \right) : i = 1,\ldots,n \right\}$ be the censored
data of $\left( \mathbf{X}^{\ast}, \mathbf{Y}^{\ast} \right)$.

The main statistical problem of this paper is to test the independence of $X_i$s and $Y_i$s in bivariate paired data using their ranks.
Let $\mathbf{R}^{X \ast} = \left( R_{1}^{X \ast}, R_{2}^{X \ast}, \ldots, R_{n}^{X \ast} \right)$ and $\mathbf{R}^{Y \ast} = \left( R_{1}^{Y \ast}, R_{2}^{Y \ast}, \ldots, R_{n}^{Y \ast} \right)$ be the ranking of complete observation among $\mathbf{X}^{\ast} = \left( X_{1}^{\ast}, \ldots, X_{n}^{\ast} \right)$ and $\mathbf{Y}^{\ast} = \left( Y_{1}^{\ast}, \ldots, Y_{n}^{\ast} \right)$, respectively.
To be specific, $R_{i}^{X \ast}$ is the rank of $X_{i}^{\ast}$ among $X_{1}^{\ast}, \ldots, X_{n}^{\ast}$, and $R_{i}^{Y \ast}$ is similarly defined.
Let $\mathbf{S}_{X}$ and $\mathbf{S}_{Y}$ be the restricted permutation space (the detailed definition is followed in Section 2.1)
that contains all available ranks $\mathbf{R}^{X}$ and $\mathbf{R}^{Y}$ given the incomplete data $\mathbf{X}$ and $\mathbf{Y}$.
For a given pair of ranks $\left( \mathbf{R}^{X}, \mathbf{R}^{Y} \right) = \left\{ \left( R_{i}^{X}, R_{i}^{Y} \right) : i = 1,\ldots,n \right\}$, the rank-based correlation, Kendall's tau $\tau = \tau \left( \mathbf{R}^{X}, \mathbf{R}^{Y} \right)$ is defined as
\begin{equation} \label{eqn:tau}
\tau\left( \mathbf{R}^{X}, \mathbf{R}^{Y} \right) = \frac{\sum_{i < j} a_{ij} b_{ij}}{n(n-1)/2},
\end{equation}
where $a_{ij} = 2 I(R_{i}^{X} < R_{j}^{X}) - 1$ and $b_{ij} = 2 I(R_{i}^{Y} < R_{j}^{Y}) - 1$, and $I (A)$ is the indicator function of
the event $A$. To test the independence of the bivariate sample, we suggest to use the simple average of $\big\{ \tau\left( \mathbf{R}^{X}, \mathbf{R}^{Y} \right) : \mathbf{R}^{X} \in \mathbf{S}_X, \mathbf{R}^{Y}
\in \mathbf{S}_Y \big\}$. To evaluate the testing statistics $\tau^{RP}$, we develop a Markov chain Monte Carlo (MCMC) procedure to sample from the uniform distribution on $\mathbf{S}_X$ and $\mathbf{S}_Y$.

The paper is organized as follows. In Section 2, we introduce a general procedure to test the independence of bivariate censored
(incomplete) data and apply it to the right-censored and interval-censored data.
In Section 3, we introduce an MCMC
procedure to sample the ranks from the uniform distribution on a
restricted permutation space. 
 We numerically compare the performance of the proposed test procedure with existing methods in terms of size and power 
 in Section 4. In Section 5, we apply our procedure to three real data examples, one with right censoring and two with interval censoring. 
 Finally, in Section 6, we conclude the paper with a brief review and additional discussions on our procedure.

\section{Testing independence of bivariate censored data}

Let ${\bf R}^{X\ast}=\big(R_1^{X\ast},R_2^{X\ast},\ldots,R_n^{X\ast}\big)$ be the rank of the observations ${\bf X}^*=\big(X_1^*,X_2^*,\ldots,X_n^*\big)$.   
 Suppose we have limited observations ${\bf X}$ on ${\bf X}^{\ast}$ and only know each possible rank sets $B_{i}^X \subset \{1,2,\ldots,n\}$ for $X_i^*$, $i=1,2,\ldots,n$.
Define the restricted permutation space by ${\bf X}$
\begin{equation} \nonumber
{\bf S}_X = \big\{ {\bf R}^X \in \Pi_n : {\bf R}^X=\big(R_1^X,R_2^X,\ldots,R_n^X \big), R_i^X \in B_i^X, i=1,2,\ldots,n \big\},
\end{equation}
where $\Pi_{n}$ is the collection of all permutations of $\{ 1,2,\ldots,n \}$.
For ${\bf Y}$, ${\bf R}^{Y\ast}$, ${\bf R}^{Y}$, $B_i^{Y}$, and ${\bf S}_Y$ are similarly defined.

In this section, 
we aim at testing  the hypothesis `$\mathcal{H}_0$: $X$ and $Y$ are independent of each other' based on the ranks of the bivariate censored data. To do it, we suggest to use
a test statistic based on $\{ \tau( \mathbf{R}^{X}, \mathbf{R}^{Y}) : \mathbf{R}^{X} \in \mathbf{S}_{X}, \mathbf{R}^{Y} \in \mathbf{S}_{Y} \}$.
Among many choices, 
we consider a test statistic, the simple average of $\tau( \mathbf{R}^{X}, \mathbf{R}^{Y})$ over
the restricted space $\mathbf{S}_{X} \times \mathbf{S}_{Y}$,
\begin{equation} \label{eqn:tau-avg}
\tau^{RP} = \frac{1}{\left\vert \mathbf{S}_{X} \right\vert \left\vert \mathbf{S}_{Y} \right\vert} \sum_{\mathbf{R}^{X} \in \mathbf{S}_{X}, \mathbf{R}^{Y} \in \mathbf{S}_{Y}} \tau\left( \mathbf{R}^{X}, \mathbf{R}^{Y} \right)
\end{equation}
where $|{\bf S}_X|$ and $|{\bf S}_Y|$ are the cardinality of ${\bf S}_X$ and ${\bf S}_Y$.
 Hereafter, we refer the proposed test statistic to the restricted permutation (RP) statistic.
We remark that the above statistic is equivalent to the original Kendall's tau statistic when the data is complete and 
its extensions to censored data are discussed by many authors in the literature 
\citep{Oakes:1982,Martin:2005,Hsieh:2010,Shen:2016}.
In practice, the evaluation of the test statistic $\tau^{RP}$
is not easy. Instead, we approximate the statistic with samples $\left\{ \left( \mathbf{R}_{b}^{X},\mathbf{R}_{b}^{Y} \right) : b = 1,\ldots,B \right\}$ from the uniform distribution on $\mathbf{S}_{X} \times \mathbf{S}_{Y}$ as
\begin{equation} \nonumber
\hat{\tau}^{RP} = \frac{1}{B} \sum_{b=1}^B \tau( \mathbf{R}_{b}^{X}, \mathbf{R}_{b}^{Y}).
\end{equation}

We approximate the null distribution of $\hat{\tau}^{RP}$ using a permutation procedure. 
We permute the indexes of $Y_i$'s to those of $Y_{\eta(i)}$'s, where $\eta=(\eta(1), \eta(2), \ldots, \eta(n)) \in \Pi_n$ and $\Pi_n$ is the collection of all permutations of $\{1,2,\ldots,n\}$. This defines the restricted permutation space ${\bf S}_{Y}(\eta)$ for each $\eta \in \Pi_n$ and its permuted ranks $\mathbf{R}^{Y} (\eta) = \left( R_{\eta(1)}^{Y}, R_{\eta(2)}^{Y}, \ldots, R_{\eta(n)}^{Y} \right)$. Suppose we have the ranks $\mathbf{R}_b^{Y}, b=1,2,\ldots,B$ from the uniform distribution on ${\bf S}_{Y}$. For a given $\eta \in \Pi_n$, $ \left\{ \mathbf{R}_b^{Y} (\eta): b=1,2,\ldots,B \right\}$ are samples from the uniform distribution on ${\bf S}_{Y} (\eta)$. Thus, for each $\eta \in \Pi_n$, we have
\begin{equation} \nonumber
\hat{\tau}^{RP}(\eta) = \frac{1}{B} \sum_{b=1}^B \tau( \mathbf{R}_{b}^{X}, \mathbf{R}_{b}^{Y} (\eta)) \nonumber\\
\end{equation}
We approximate the null distribution of $\tau^{RP}$ with $\big\{\hat{\tau}^{RP}(\eta): \eta \in \Pi_n \big\}$.

It is worth noting that the proposed RP statistic is only valid for testing independence,
since the RP statistic $\hat{\tau}^{RP}$ is an approximate of the uniform average of all possible $\tau({\bf R}^X,
{\bf R}^Y)$ for $({\bf R}^X, {\bf R}^Y) \in {\bf S}_X \times {\bf S}_Y$. The conditional distribution of $\tau({\bf R}^X,
{\bf R}^Y)$ given the incomplete data is generally not uniform, and this makes $\tau^{RP}$ be biased in estimating
the true parameter $\tau$. We will numerically understand this in Section 4.

In the following subsections, we characterize the restricted permutation space for
two types of bivariate censored data, bivariate right- and interval-censored data.
We remark again that
the proposed testing procedure is not restricted to bivariate right- and interval-censored data
as we know from the definition of the RP statistic. The RP statistic can be used for testing the independence of 
any bivariate incomplete data where the restricted permutation space of ranks is well-defined.

\subsection{Bivariate right-censored data}

Right-censored data is commonly encountered
when we analyze survival times since
some participants leave the study, or the study ends before a primary outcome occurs \citep{Cuzick:1982,Oakes:1982,Pan:2018}.

Bivariate right-censored data consist of
paired survival times $\left((X_i, \delta^X_i), (Y_i,\delta^Y_i)\right)$ for
$i=1,2,\ldots, n$, where $X_i$ and $Y_i$ denote the
survival times or censoring times of two targeted outcomes for the $i$th subject, $\delta^X_i$ is an indicator of
censoring on $X_i$ whose value is zero if censoring occurs on $X_i$ and 1 otherwise, and $\delta^Y_i$ is similarly defined as $\delta^X_i$.
To be more specific, suppose $(X_i^*, Y_i^*)$ and $(X_i^c, Y_i^c)$ are paired survival times and censoring
times of the $i$th subject, respectively. We observe $X_i =\min(X_i^*, X_i^c)$ and $Y_i= \min(Y_i^*, Y_i^c)$,
and record $\left( (\min(X_i^*, X_i^c), I(X_i^* \le X_i^c)), (\min(Y_i^*, Y_i^c), I(Y_i^* \le Y_i^c))\right)$. We let
$\delta^X_i= I(X_i^* \le X_i^c)$ and $\delta^Y_i= I(Y_i^* \le Y_i^c)$.

Suppose we observe
$\big\{ (x_i,\delta^X_i),~i=1,2,\ldots,n\big\}$ and the ranks of tied observations are exchangeable. Then, the
restricted permutation space of ranks of ${\bf X}$, ${\bf S}_{\bf X}$ is defined as in the following proposition. Further,
under the same assumption to the observations $\big\{ (y_i,\delta^Y_i),~i=1,2,\ldots,n\big\}$, the restricted
permutation space of ranks ${\bf S}_{\bf Y}$ is similarly defined.

\begin{proposition} \label{proposition:right_rank}
Let $\{I_U, I_R\}$ be a partition of an index set $\{1,2,\ldots,n\}$ corresponding
to the uncensored and
right-censored observations such that $(x_i, \delta^X_i)
= (x_{i}^*,1)$ for $k\in I_U$ and $(x_i,\delta^X_i) =
(x_{i}^c, 0)$ for $k\in I_R$,
where $x_{i}^*$s and $x_{i}^c$s are observed survival times
of uncensored and right-censored data, respectively.
Then, the set $B_i^X$ of possible ranks of the $i$th observation
given the observation $\big\{ (x_i, \delta^X_i), i=1,2,\ldots,n \big\}$ is
defined as
\[
B_i^X=\left\{\begin{array}{lcl}
\big\{
R : \sum_{j\in I_U} I(x_j<x_i) + 1 \le R \le
\sum_{j\in I_U} I(x_j\le x_i) + \sum_{j\in I_R} I(x_j < x_i) \big\}
&\mbox{ for }& i\in I_U,\\[2mm]
\big\{
R : \sum_{j\in I_U} I(x_j\le x_i) + 1 \le R \le n \big\}
&\mbox{ for }& i \in I_R.
\end{array}
\right.
\]
With $B_i^X$s above, the restricted permutation space ${\bf S}_{\bf X}$ is defined as
\[
\bS_X = \{\bR^X \in \Pi_n : \bR^X = (R^X_1, R^X_2,\ldots, R^X_n), R^X_i \in B^X_i, i=1,2,\ldots, n\}.
\]
\end{proposition}

\subsection{Bivariate (case-2) interval-censored data}

Interval-censored data arises when survival time $X$ is discretely observed at a sequence of examination times. If the
number of examination times is $k$, say $T_1, T_2, \ldots, T_k$, we observe the survival time $X$ is in one
of the interval $(T_{j-1}, T_j]$, $j=1,2,\ldots,k+1$, with $t_0=0$ and $T_{k+1}=\infty$, and we call the time $X$
is `case-k interval-censored' \citep{Wellner:1995}. 
Further, we could find that, for $k \ge 3$, every case-k interval-censored data can
be written as case-2 interval-censored data without information loss. In this paper, we especially focus on the case-2 interval
censoring \citep{Wellner:1995,Sun:2006}.

In the case-2 interval-censored data,
each subject of the study is observed twice at
given times $T_1$ and $T_2$ $(T_1 < T_2)$.
Hence, we only know the information that
each survival time is contained in one of three intervals
$(0,T_1]$, $(T_1, T_2]$, and $(T_2, \infty)$.
To be specific,
the $i$th observation with observation times $T_{i1}$ and $T_{i2}$ $(T_{i1} < T_{i2})$ can be represented as $(T_{i1}, T_{i2}, \delta_{i 1}, \delta_{i2})
= (T_{i1}, T_{i2}, I(X_i \le T_{i1}), I(T_{i1} < X_i \le T_{i2}))$,
where $(T_{i1},T_{i2},0,0)$ denotes $X_i \in (T_{i2}, \infty)$.
In addition, the case-2 interval-censored observations can also be written as
$\{ (L_i, R_i]\}$, where $(L_i, R_i] \in \big\{(0, T_{i1}], (T_{i1}, T_{i2}],
(T_{i2}, \infty) \big\}$ with $0 < T_{i1} \le T_{i2} < \infty$. Suppose $\big\{ (L_i,R_i], ~ i=1,2,\ldots,n \big\}$ are
the observed case-2 interval-censored data for ${\bf X}=\{X_i, i=1,2,\ldots,n\}$. Then, the
restricted permutation space of ranks of ${\bf X}$, ${\bf S}_{\bf X}$ is defined as in the following proposition. Further,
that for ${\bf Y}$, ${\bf S}_{\bf Y}$ is similarly defined.

\begin{proposition} \label{proposition:case2_rank}
Let
$\{I_L, I_C, I_R, I_U\}$ be the partition of the index set $\{1,2,\ldots,n\}$ corresponding
to the left-censoring, right-censoring, interval-censoring and uncensoring
of each observation. To be specific, $(L_i, R_i] = (0, T_{i1}]$ for $i\in I_L$,
$(L_i, R_i] = (T_{i1}, T_{i2}]$ for $i \in I_C$,
$(L_i, R_i] = (T_{i2}, \infty)$ for $i \in I_R$,
and $(L_i, R_i] = \{ X_i \}$ for $i\in I_U$. Suppose we again assume
the ranks of tied observations are exchangeable.
Then, the set $B_i^X$ of possible ranks of the $k$th complete observation $X_i$
given the censored observation $\big\{ (L_i,R_i], ~ i=1,2,\ldots,n \big\}$ is
defined as
\[
B_i^X = \left\{\begin{array}{lcl}
\big\{ R : 1 \le R \le |I_L|+\sum_{j \in I_R }^n I(L_j < R_i) +
\sum_{j \in I_U }^n I(L_j < R_i) + \sum_{j \in I_C} I(L_j \le R_i) \big\}
&\mbox{ for }& i \in I_L,\\[2mm]
\big \{ R : \sum_{j\notin I_R} I(R_j \le L_i) < R \le
|I_L|+\sum_{j\in I_R \cup I_C} I(L_j<R_i) + \sum_{j\in I_U} I(L_j \le R_i)
\big\} &\mbox{ for }& i \in I_C,\\[2mm]
\big\{ R : \sum_{j\notin I_R} I(R_j \le L_i) < R \le n \big\}
&\mbox{ for }& i\in I_R,\\[2mm]
\big\{ R : \sum_{j\notin I_R} I(R_j < L_i) < R \le
|I_L|+\sum_{j\in I_R \cup I_C} I(L_j<R_i) + \sum_{j\in I_U} I(L_j \le R_i)
\big\} &\mbox{ for }& i \in I_U.
\end{array}
\right.
\]
With $B_i^X$s above, the restricted permutation space ${\bf S}_{\bf X}$ is defined as
\[
\bS_X =\big\{\bR^X \in \Pi_{n} : \bR^X = (R^X_1, R^X_2,\ldots, R^X_n), R^X_i \in B^X_i, i=1,2,\ldots, n \big\}.
\]
\end{proposition}

\section{MCMC procedure for uniform sample on ${\bf S}$} 

In Section 2, we propose the RP test statistic calculated by 
the samples from the uniform distribution on ${\bf S}_X \times {\bf S}_Y$.
Since the sample $({\bf R}^X, {\bf R}^Y) \in {\bf S}_X \times {\bf S}_Y$ can
be obtained by the independent random samples ${\bf R}^X \in {\bf S}_X$ and ${\bf R}^{Y} \in {\bf S}_Y$, in this section, we focus on generating a random sample from the uniform distribution 
on the restricted permutation space ${\bf S}_X$, which we call `uniform sample ${\bf r}$ from the space ${\bf S}_X$'.
For notational simplicity,
we denote ${\bf S}_X$ and ${\bf R}^X$ as ${\bf S}$ and ${\bf R}$, respectively.

Suppose we consider a graph $G=\big({\bf S},{\bf E} \big)$, where the edge sets are defined as the following exchanging operations:
two nodes ${\bf R}_1=\big( R_{11}, R_{12},\ldots,R_{1n} \big)$ and ${\bf R}_2=\big(R_{21},R_{22},\ldots,R_{2n} \big)$ in
${\bf S}$ are connected if there exists a pair $(c,d)$ where $c,d \in \{1,2,\ldots,n\}, c \neq d$ such that
$R_{1c}=R_{2d}$, $R_{1d}=R_{2c}$ and $R_{1k}=R_{2k}$ for all $k \neq c,d$. To be specific, let $e\big( {\bf R}_1, {\bf R}_2\big)$ be an indicator function of edge connection
between two nodes ${\bf R}_1$ and ${\bf R}_2$ (i.e.,
$e\big({\bf R}_1, {\bf R}_2 \big) =1$ if ${\bf R}_1$ and ${\bf R}_2$ are connected, 0 otherwise). The graph of the restricted permutation
space $G = \big({\bf S},{\bf E} \big)$ is defined with the edge set ${\bf E} =\big\{ (i,j) : e \big({\bf R}_i, {\bf R}_j \big) =1, {\bf R}_i, {\bf R}_j \in {\bf S} \big\}$.
The degree of each node ${\bf R}$ is
${\rm deg} \big( {\bf R}\big) = \big| \big\{ {\bf R}^{\prime} \in {\bf S} : e \big({\bf R}, {\bf R}^{\prime} \big)=1 \big\} \big|$. We 
refer readers to \citet{Roselle:1974}, \citet{Diaconis:2001}, \citet{Lim:2006} and their references for 
more details on restricted permutation space.

To generate a random sample from the uniform distribution on ${\bf S}$, we first define two random walks, 
$\{ V_{t} \}_{t=1}^{\infty}$ and $\{ W_{t} \}_{t=1}^{\infty}$, on the space $G=(\mathbf{S},\mathbf{E})$ in the following.

\begin{algorithm} \label{alg:target}
Given that $V_{t} = {\bf R}=(R_1,R_2,\ldots,R_n) \in {\bf S}$, $V_{t+1}$ is generated as follows :
\begin{enumerate}
\item[(P1)] Randomly choose $c,d \in \{ 1,2,\cdots,n\} (c \neq d)$ and define $\tilde{\bf R}=(\tilde{R}_1, \tilde{R}_2,\ldots,\tilde{R}_n) \in \Pi_{n}$ as $\tilde{R}_{c} = R_{d}$, $\tilde{R}_{d} = R_{c}$, and $\tilde{R}_{k} = R_{k}$ for $k \neq c,d$.
\item[(P2)] If $\tilde{\bf R} \nin {\bf S}$, then {\it stay at ${\bf R}$}. \ie $V_{t+1} = {\bf R}$.
\item[(P3)] If $\tilde{\bf R} \in {\bf S}$, then move to $\tilde{
\bf R}$. \ie $V_{t+1} = \tilde{\bf R}$.
\end{enumerate}
\end{algorithm}

\begin{algorithm} \label{alg:add}
Given that $W_{t} = {\bf R}=(R_1,R_2,\ldots,R_n) \in {\bf S}$, $W_{t+1}$ is generated as follows :
\begin{enumerate}
\item[(Q1)] Randomly choose $c,d \in \{ 1,2,\cdots,n\} (c \neq d)$ and define $\tilde{\bf R} = (\tilde{R}_1, \tilde{R}_2,\ldots,\tilde{R}_n) \in \Pi_{n}$ as $\tilde{R}_{c} = R_{d}$, $\tilde{R}_{d} = R_{c}$, and $\tilde{R}_{k} = R_{k}$ for $k \neq c,d$.
\item[(Q2)] If $\tilde{\bf R} \nin {\bf S}$, then {\it repeat} (Q1).
\item[(Q3)] If $\tilde{\bf R} \in {\bf S}$, then move to $\tilde{\bf R}$. \ie $W_{t+1} = \tilde{\bf R}$.
\end{enumerate}
\end{algorithm}

The following theorem states the limiting distribution of two random walk.

\begin{theorem}\label{thm1}
If the graph $G=(\mathbf{S},\mathbf{E})$ is connected and non-bipartite, two random walks $\{ V_{t} \}_{t=1}^{\infty}$ and $\{ W_{t} \}_{t=1}^{\infty}$ defined by Algorithm \ref{alg:add} and \ref{alg:target}, respectively, are ergodic, which implies that each of two random walks has the limiting distribution. If we let $\mathbf{p}$ and $\mathbf{q}$ denote the limiting distributions of random walks $\{ V_{t} \}_{t=1}^{\infty}$ and $\{ W_{t} \}_{t=1}^{\infty}$, then $\mathbf{p}(\mathbf{R}) \propto 1$ and $\mathbf{q}(\mathbf{R}) \propto \mathrm{deg}(\mathbf{R})$.
\end{theorem}

\begin{proof}
According to \citet{Chung:1997}, we can easily see that two Markov chains $\{ V_{t} \}_{t=1}^{\infty}$ and $\{ W_{t} \}_{t=1}^{\infty}$ are ergodic. In this proof, we calculate their limiting distributions. Let ${\rm P} = ({\rm P}_{uv}), {\rm Q}= ({\rm Q}_{uv}) \in \mathbb{R}^{\vert {\bf S} \vert \times \vert {\bf S} \vert}$ be the transition matrix of two Markov chains, respectively. Given $V_{t} = u \in \bS$ and $W_{t} = u \in \bS$, we need to calculate the probability of the event $\{ V_{t+1} = v \}$ and $\{ W_{t+1} = v \}$ for each $v \in \bS$, respectively. Note that by the definition of the connection of two nodes, for each node $v\in \bS$ with $e(u,v) = 1$, $\{ V_{t+1} = v \}$ has the same probability. The number of such nodes is $\deg(u)$. There are $n(n-1)/2$ unordered pairs $\{c,d\}$ ($c \neq d$) that can be chosen in the step (P1) in Algorithm \ref{alg:target}. Among them, $n(n-1)/2 - \deg(u)$ pairs result in $V_{t+1} = u$, and the other $\deg(u)$ pairs make $V_{t+1}$ move to other node that corresponds to the chosen unordered pair. Hence, we obtain
\begin{equation} \nonumber
{\rm P}_{uv} = \Pr \left( V_{t+1} = v | V_{t} = u \right) = \left\{
\begin{array}{c l}
1 - \frac{{\rm deg}(u)}{K} & \mbox{if } v=u, \\
\frac{1}{K} & \mbox{if } v \neq u \mbox{ and } e(v,u)=1, \\
0 & \mbox{if } v \neq u \mbox{ and } e(v,u)=0,
\end{array}
\right.
\end{equation}
where $K = n(n-1)/2$. Similarly, for each node $v\in \bS$ with $e(u,v) = 1$, $\{ W_{t+1} = v \}$ has the same probability. However, in this case, there is no chance to move to nodes that are not connected to $u$. Note that in Algorithm \ref{alg:add}, we uniformly choose a node among the nodes connected to the node $u$. Thus, we have
\begin{equation} \nonumber
{\rm Q}_{uv} = \Pr \left( W_{t+1} = v | W_{t} = u \right) = \frac{e(u,v)}{{\rm deg}(u)}.
\end{equation}

Now, we let $\mathbf{p}$ and $\mathbf{q}$ denote the limiting distributions of random walks $\{ V_{t} \}_{t=1}^{\infty}$ and $\{ W_{t} \}_{t=1}^{\infty}$. Then, by the results about random walks in \citet{Chung:1997}, $\textbf{p}$ and $\textbf{q}$ are proportional to a vector $(\sum_{v}w(u,v) : u \in \bS)$ of which each element is the sum of the weights $w(u,v)$. If we let $w_{\rm P}$ and $w_{\rm Q}$ be the weights corresponding to Algorithm \ref{alg:target} and \ref{alg:add}, respectively, then it is easy to check that they are given as follows.
\begin{eqnarray}
w_{\rm P}(u,v) & = & \left\{
\begin{array}{c l}
K - \sum_{w}e(u,w) & \mbox{if } v=u, \\
1 & \mbox{if } v \neq u \mbox{ and } e(v,u)=1, \\
0 & \mbox{if } v \neq u \mbox{ and } e(v,u)=0,
\end{array}
\right. \nonumber\\
w_{\rm Q}(u,v) & = & e(u,v). \nonumber
\end{eqnarray}
Therefore, we have $\textbf{p}(u) \propto \sum_{v} w_{\rm P}(u,v) = K$ and $\textbf{q}(u) \propto \sum_{v} w_{\rm Q}(u,v) = \deg(u)$, which completes the proof.
\end{proof}

Since the limiting distribution of the Markov chain $\{ V_{t} \}_{t=1}^{\infty}$ is uniform on $\bf{S}$, we can utilize the chain to generate uniform samples on $\bf{S}$. However, the Markov chain $\{ V_{t} \}_{t=1}^{\infty}$ could be mixed slowly compared to $\{ W_{t} \}_{t=1}^{\infty}$, if the structure of the graph $G = (\mathbf{S},\mathbf{E})$ is sparse. To accelerate the mixing, we adapt the idea of parallel tempering \citep{Geyer:1995}. In the following algorithm, we generate a random walk $\{ (V_{t}^{c},W_{t}^{c}) \}_{t=1}^{\infty}$ on $\bf{S} \times \bf{S}$ by coupling two chains $\{ V_{t} \}_{t=1}^{\infty}$ and $\{ W_{t} \}_{t=1}^{\infty}$.

\begin{algorithm} \label{alg:coupled}
Given that $(V_{t}^{c},W_{t}^{c}) = (\mathbf{R}_{01},\mathbf{R}_{02})$, $(V_{t+1}^{c},W_{t+1}^{c})$ is generated as follows :
\begin{itemize}
\item[(C1)] Put $(V_{t},W_{t}) = (\mathbf{R}_{01},\mathbf{R}_{02})$, and generate $V_{t+1}$ and $W_{t+1}$ using Algorithm \ref{alg:target} and \ref{alg:add}, respectively. Write the generated state as $(V_{t+1},W_{t+1}) = (\mathbf{R}_{1},\mathbf{R}_{2})$.
\item[(C2)] Denote the acceptance probability by
\begin{equation} \nonumber
\kappa = \kappa(\mathbf{R}_{1},\mathbf{R}_{2}) = \min \left\{1, \frac{{\bf p}({\bf R}_{2}) {\bf q}({\bf R}_{1})}{{\bf p}({\bf R}_{1}) {\bf q}({\bf R}_{2})} \right\} = \min\left\{ 1, \frac{\deg({\bf R}_{1})}{\deg({\bf R}_{2})} \right\}.
\end{equation}
\item[(C3)] We accept the transition with probability $\kappa$, and reject it with probability $1-\kappa$. That is, we obtain $(V_{t+1}^{c},W_{t+1}^{c}) = (\mathbf{R}_{2},\mathbf{R}_{1})$ with probability $\kappa$, and $(V_{t+1}^{c},W_{t+1}^{c}) = (\mathbf{R}_{1},\mathbf{R}_{2})$ with probability $1-\kappa$
\end{itemize}
\end{algorithm}
Note that the steps (C2) and (C3) are equivalent to the following steps (C2)$^{\prime}$ and (C3)$^{\prime}$.
\emph{\begin{itemize}
\item[(C2)$^{\prime}$] Generate a random variable $U$ from uniform distribution on $[0,1]$.
\item[(C3)$^{\prime}$] We obtain $(V_{t+1}^{c},W_{t+1}^{c})$ as follows.
\begin{equation} \nonumber
(V_{t+1}^{c},W_{t+1}^{c}) = \left\{ \begin{array}{c l}
(\mathbf{R}_{2},\mathbf{R}_{1}) & \mbox{if } U \le \frac{\deg({\bf R}_{1})}{\deg({\bf R}_{2})}, \\
(\mathbf{R}_{1},\mathbf{R}_{2}) & \mbox{if } U > \frac{\deg({\bf R}_{1})}{\deg({\bf R}_{2})}.
\end{array} \right.
\end{equation}
\end{itemize}}

The coupled chain generated by the above algorithm is known as Metropolis-Hastings coupled Markov chain. Due to the result of \citet{Geyer:1995}, we obtain the following theorem.

\begin{theorem} \label{thm2}
Under the assumption presented in Theorem \ref{thm1}, the coupled Markov chain $\{ (V_{t}^{c},W_{t}^{c}) \}_{t=1}^{\infty}$ has the same limiting distribution as $\{ (V_{t},W_{t}) \}_{t=1}^{\infty}$.
\end{theorem}

We numerically confirm that the parallel tempering accelerates the convergence of the distribution of the target chain to the uniform distribution. In this study, we compare the convergence speed of the target chain distribution with and without the parallel tempering.
Four types of graphs with 100 nodes are used in simulation: i) circulant graph $C_{101}^{1}$ with one jump (Here, the number of nodes is 101 because the graph should not be bipartite.), ii) circulant graph $C_{100}^{1,2,3}$ with three jumps, iii) random 3-regular graph, and iv) the graph generated by the Watts-Strogatz small-world model. First, we generate two pairs of Markov chains on each graph. One is composed of two independent chains with transition probabilities ${\rm P}$ and ${\rm Q}$. The other consists of two chains with transition probabilities ${\rm P}$ and ${\rm Q}$ that are Metropolis-coupled. To remove the effect of the initial value, we discard the first 5,000 samples. Then, we take only one sample per 100 samples to reduce the correlation between neighboring samples. Finally, we calculate the $L_{\infty}$-distance and the total variation distance between the distribution of the target chain and the uniform distribution. The two metrics are defined as follows : for distributions $f,g : V \rightarrow [0,1]$ on the set $V = \{ 1,2, \ldots, n \}$ of nodes, the $L_{\infty}$-distance is defined as
$d_{\rm sup}(f,g) = \max_{v \in V}\left\vert f(v) - g(v) \right\vert,
$
and the total variation distance is defined as
$d_{\rm tv}(f,g) = \max_{A \subseteq V}\left\vert f(A) - g(A) \right\vert = \frac{1}{2} \sum_{v \in V}\left\vert f(v) - g(v) \right\vert.
$
The results are reported in Appendix B.

\section{Numerical study}

In this section, we numerically investigate the sizes and powers of
the proposed independence test procedure based on the RP statistics
for bivariate right- and interval-censored data.
To generate correlated bivariate survival times, we
consider Clayton and Frank copula models  as in \citet{Kim:2015}.
To be specific, let $F(x,y)$ be a joint cumulative distribution function of complete
observation $(X^*, Y^*)$, and $F_X(x)$ and $F_Y(y)$ be marginal
cumulative distribution functions of $X^*$ and $Y^*$, respectively.
The Clayton copula model \citep{Clayton:1978} is defined as
\[
F(x,y) = \big( (F_X(x))^{-\alpha} + (F_Y(y))^{-\alpha} -1 \big)^{-1/\alpha},
\]
where $\alpha = 2\tau/(1-\tau) > 0$ and $\tau$ denotes the Kendall's $\tau$.
The Frank copula model \citep{Frank:1979} is defined as
\[
F(x,y) = -\frac{1}{\beta} \log \left(
1 + \frac{ (e^{-\beta F_X(x)} -1)(e^{-\beta F_Y(y)}-1)}{e^{-\beta} -1}
\right),
\]
where $\beta >0$ controls dependency of $X^*$ and $Y^*$. It is known that
$\tau = 1 + 4\beta^{-1} ( D_1(\beta) -1 )$, where $D_1(\beta) = \beta^{-1}\int_{0}^{\beta} t(e^t -1)^{-1} ~dt$ is the first-order Debye function.
With the given copula model, we first generate $(U_1,U_2)$ from the given copula model and apply the inverse transformation of marginal distribution functions $(F_X^{-1}(U_1), F_Y^{-1}(U_2))$.
For marginal distributions of $X^*$ and $Y^*$, we consider $F_X(x) = 1-e^{-0.1x}$
and $F_Y(y) = 1-e^{-0.1y}$ following \citet{Kim:2015}.
In this study, the bivariate survival times $\{(X_i^*, Y_i^*)\}_{i=1}^n$ are generated with
$\tau = 0,~1/5,~1/3$ and $n=50,~ 100$, where
$\tau = 1/5,~1/3$ are corresponding to $\alpha=1/2,~ 1$ for Clayton copula
and $\beta \approx 1.8609,~3.3058$ for Frank copula, respectively.
We evaluate the sizes and powers based on 500 simulations, where
both the number of permutation for approximating null distribution and
the number of MC samples from the restricted permutation space
are set as 1000.

\subsection{Bivariate right-censored data}

To generate right-censored data, we assume that the censoring times $X_i^c$ and $Y_i^c$ are 
independent with the event times $X_i^*$ and $Y_i^*$ and
follows the uniform distribution $U(0, c_R)$.
With the generated complete bivariate observations $\{ (X_i^*, Y_i^*), i=1,2,\ldots,n\}$
and censoring times $\{ (X_i^c, Y_i^c), i=1,2,\ldots,n\}$,
we define the right-censored data $\{ (X_i, \delta_i^X), (Y_i, \delta_i^Y), i=1,2,\ldots,n\}$
with $X_i = \min(X_i^*, X_i^c)$, $\delta_i^X=I(X_i^*\le X_i^c)$,
$Y_i = \min(Y_i^*, Y_i^c)$, and $\delta_i^Y=I(Y_i^*\le Y_i^c)$.
We consider $c_R=9,~15$,
where censoring rates with $c_R=9,~15$ are 65.8\% and 51.8\%, respectively.

For testing the independence of the bivariate right-censored data,
several methods are introduced in the literature.
Among the existing methods, for the purpose of comparison,
we consider two testing procedures using Kendall's tau
proposed by \citet{Oakes:1982} and \cite{Kim:2015}.
We denote the Kendall's tau defined in \cite{Oakes:1982}
and \citet{Kim:2015} as $\hat{\tau}^{O}$ and $\hat{\tau}^{NP}$, respectively.
In \citet{Oakes:1982}, the Kendall's tau $\hat{\tau}^{\rm O}$ for the right-censored data
is defined as
\begin{equation} \label{tau:oakes}
\hat{\tau}^{O} = \frac{\sum_{i<j} a_{ij} b_{ij}}{n(n-1)/2},
\end{equation}
where $a_{ij} = 1$ if $X_i > X_j$ and $(\delta_i^X, \delta_j^X) \in
\{ (0,1),(1,1)\}$, $-1$ if $X_i < X_j$ and $(\delta_i^X, \delta_j^X) \in
\{ (1,0),(1,1)\}$, and $0$ otherwise, and $b_{ij}$ is similarly defined
with $Y_i$ and $Y_j$. In \cite{Kim:2015}, the Kendall's tau $\hat{\tau}^{NP}$ is defined
as
\begin{equation} \label{tau:score}
\hat{\tau}^{NP} = \frac{\sum_{i,j} \tilde{a}_{ij} \tilde{b}_{ij}}{
\sqrt{(\sum_{i,j} \tilde{a}_{ij}^2)(\sum_{i,j} \tilde{b}_{ij}^2)}},
\end{equation}
where $\tilde{a}_{ij} =2 P\big(X_i^* > X_j^* ~|~ (X_i, \delta_i^X), (X_j, \delta_j^X)\big)-1$ and $\tilde{b}_{ij} = 2P\big(Y_i^* > Y_j^* ~|~ (Y_i, \delta_i^Y), (Y_j, \delta_j^Y)\big)-1$. The conditional probabilities $\{\tilde{a}_{ij}\}$ and $\{\tilde{b}_{ij}\}$ are estimated by
the nonparametric maximum likelihood estimators (NPMLEs) of the marginal distribution functions $F_X(x)$ and $F_Y(y)$. In this paper, we estimate the marginal distribution functions $F_X(x)$ and $F_Y(y)$ with the R function
\texttt{icfit} in R package \texttt{interval}.
The test statistics $Z_{O}= \hat{\tau}^{ O}/\sqrt{{\rm Var}(\hat{\tau}^{O})}$
and $Z_{NP} = \hat{\tau}^{NP}/\sqrt{{\rm Var}(\hat{\tau}^{NP})}$
are used for testing independence, where their asymptotic
null distribution is the standard normal distribution.

To compare fairly, we evaluate p-values of $\hat{\tau}^{O}$, $\hat{\tau}^{NP}$ and $\hat{\tau}^{RP}$ with the approximated
null distribution by permutation as well as their asymptotic null distributions. The RP statistic $\hat{\tau}^{RP}$
is the mean of $\{\tau(R_{b}^X, R_{b}^Y): b=1,\ldots,B\}$ and then
$Z_{RP} = \hat{\tau}^{RP}/\sqrt{{\rm Var}(\hat{\tau}^{RP})} \stackrel{d}{\to} N(0,1)$
by the central limit theorem. We estimate variances of
$\hat{\tau}^O$, $\hat{\tau}^{NP}$, and $\hat{\tau}^{RP}$ by the permutation and plug in the variance estimates in $Z_O$, $Z_{NP}$, and $Z_{RP}$. We report the bias ($(1/m)\sum_{i=1}^m \hat{\tau}_i - \tau$),
the mean squared error (MSE, $(1/m)\sum_{i=1}^m (\hat{\tau}_i - \tau)^2$),
empirical powers evaluated by asymptotic null distribution (EP-A) and
empirical powers evaluated by permutation (EP-P) in Table 1, where $m$ is the number of simulations, EP-A and EP-P are calculated by
the ratios of the number of rejections to
the number of simulations with p-values from the asymptotic null
distribution and the approximated null distribution by permutation,
respectively.

From Table 1, we observed several features.
First, the asymptotic null distributions and approximated null distribution by permutation are similar and well-approximated to the true null distribution
since the sizes of independence test of $\hat{\tau}^O$, $\hat{\tau}^{NP}$, and
$\hat{\tau}^{RP}$ are well-controlled with the given level $0.05$, and all EP-A and EP-P values are similar.
Second, powers of all three methods increase when the censoring rates
decrease from 65.8\% ($c_R=9$) to 51.8\% ($c_R=15$) except the case of $(n=50, \tau=1/5)$.
In the case $(n=50, \tau=1/5)$, the decrements of the powers of all three methods from $c_R=9$ to $c_R=15$ are small. For instance, the decrement of
powers of $\hat{\tau}^O$, $\hat{\tau}^{NP}$, and $\hat{\tau}^{RP}$ are $0.04$, $0.012$, $0.032$ in EP-P, respectively.
Third, the biases and MSEs of all three methods increases
as the true $\tau$ increases but those of $\hat{\tau}^{NP}$ are preserved
at lower levels than those of $\hat{\tau}^{O}$ and $\hat{\tau}^{RP}$.
For example, for the case of $(n=50, \tau=1/3, c_R=15)$, the biases of $\hat{\tau}^{O}$, $\hat{\tau}^{NP}$, and $\hat{\tau}^{RP}$ are $-0.1388$, $0.0315$, and $-0.1201$, respectively.
The increment of bias of $\hat{\tau}^O$ is caused by a fixed denominator
$n(n-1)/2$ since the summation of the scores in the numerator is only
on the comparable pairs, where the number of comparable pairs can be much smaller than $n(n-1)/2$.
For $\hat{\tau}^{RP}$,
the increment of bias of $\hat{\tau}^{RP}$ is caused by the characteristic that $\hat{\tau}^{RP}$ under-estimates $\tau$
when $\tau\neq 0$ as we mentioned in the previous section.
The bias of $\hat{\tau}^{RP}$ is the smallest among three methods when $\tau=0$.
Finally, in view of powers of testing independence, no method outperforms others for all cases.
Under Clayton copula, the powers of $\hat{\tau}^O$ is greater than
those of the other two methods, and the power of $\hat{\tau}^{RP}$ is
greater than $\hat{\tau}^{NP}$. Under Frank copula, the powers of $\hat{\tau}^{NP}$ is
often similar to or larger than those of $\hat{\tau}^O$ and $\hat{\tau}^{RP}$. For this case,
our numerical study recommends the use $\hat{\tau}^O$, which is
the best under Clayton copula and the second-best under Frank copula. However, the powers of all three methods
are not significantly different.

\subsection{Bivariate interval-censored data}

Bivariate case-2 interval-censored observations are generated by three steps
as in generating bivariate right-censored observations.
First, we generate complete bivariate observations $\{(X_i^*, Y_i^*)\}_{i=1}^n$
based on the given copula model and the inverse transformation of marginal
distribution function.
Second, under the non-informative censoring assumption,
we independently generate observation times $(T_{i1}^X, T_{i2}^X)$
and $(T_{i1}^Y, T_{i2}^Y)$.
Finally, the bivariate interval-censored observations are
defined as $\big\{ \big( (T_{i1}^X, T_{i2}^X,\delta_{i1}^X, \delta_{i2}^X),
(T_{i1}^Y, T_{i2}^Y,\delta_{i1}^Y, \delta_{i2}^Y) \big)\big\}$,
where $\delta_{i1}^X = I(X_i^* \le T_{i1}^X)$,
$\delta_{i2}^X = I(T_{i1}^X < X_i^* \le T_{i2}^X)$,
and $\delta_{i1}^Y$ and $\delta_{i2}^Y$ are similarly defined.
For observation times $(T_1, T_2)$, we consider two scenarios:
\begin{itemize}
\item[]({\bf C1}): Observation times are from uniform distribution, which
assumes that the observation times are not pre-scheduled.
In this scenario,
we generate $T_{i1}$ and $Q_i$ independently from $U(0, c_L)$ and $U(0, c_R)$,
respectively. Then, we obtain $T_{i2} = T_{i1} + Q_i$ for $i=1,\ldots,n$.
We consider $c_L = 3$ and $c_R = 6, 12$, where
the rates of left, right, and interval censoring are
(13.5\%, 65\% 21.5\%) and (13.5\%, 50.5\%, 36\%) for
$c_R = 6, 12$, respectively.

\item[]({\bf C2}): Observation times are pre-scheduled. In this scenario, 
we assume that observation is conducted every three month
and a total observation period is $c_R$.
We further assume an ideal case in the sense that
all patients are observed at given observation times
and stop when the event occurs.
Thus, $(T_{i1},T_{i2}) \in \big\{ (0,3], (3,6], \ldots, (c_R,\infty)\}$.
We consider $c_R = 9, 15$,
where the rates of right, and interval censoring are
(41\%, 59\%) and (22.5\%, 77.5\%) for
$c_R = 9, 15$, respectively.

\end{itemize}

For the independence test of the bivariate interval-censored data,
we consider two existing procedures using Kendall's tau proposed by
\cite{Betensky:1999b} and \cite{Kim:2015}.
Note that we exclude the two-stage estimation procedure proposed
by \cite{SunW:2006} in the comparison of testing independence
of the bivariate interval-censored data following the comment
in \cite{Kim:2015}
that the two-stage estimation is not valid for testing independence.
As in the right-censored data,
we denote the estimator and the method of \cite{Kim:2015} as $\hat{\tau}^{NP}$.
Here, we omit the details of $\hat{\tau}^{NP}$
since basic idea and notations are the same as the ones in the bivariate right-censored
data. We refer to Section 3.2 in \cite{Kim:2015} for the estimation of conditional
probability in $\tilde{a}_{ij}$ and $\tilde{b}_{ij}$.
\cite{Betensky:1999b} extend Kendall's tau to the bivariate
interval-censored data based on the NPMLE of the joint
survival function $S(x,y)$ and the multiple imputations.
We use a notation $\hat{\tau}^{MI}$ to denote the estimator and the method by \cite{Betensky:1999b}.
To be specific, let
$\big\{ \big( (L_i^X, R_i^X], (L_i^Y, R_i^Y) \big), i=1,2,\ldots,n \big\}$ be
interval-censored observations.
In the procedure $\hat{\tau}^{MI}$, the Kendall's tau $\hat{\tau}^{MI}$ is
estimated by the following three steps.
In the first step, the joint survival function $S(x,y)$ is
estimated by the NPMLE.
Second step generates $M$ datasets where the interval-censored observations
of $\big\{ \big( (L_i^X, R_i^X], (L_i^Y, R_i^Y) \big), i=1,2,\ldots,n \big\}$ are imputed
by $\big((L_{i,(k)}^{X}, R_{i,(k)}^{X}], (L_{i,(k)}^{Y}, R_{i,(k)}^{Y}]\big) $ for $k=1,\ldots, M$ uniformly generated from
the conditional survival distribution of $\big((L_{i,(k)}^{X}, R_{i,(k)}^{X}], (L_{i,(k)}^{Y}, R_{i,(k)}^{Y}]\big) $
given $(L_{i,(k)}^{X}, R_{i,(k)}^{X}]\times (L_{i,(k)}^{Y}, R_{i,(k)}^{Y}] \subset (L_i^X, R_i^X]\times(L_i^Y, R_i^Y)$.
In the third step, the estimator $\hat{\tau}^{MI}$ is obtained by
\[
\hat{\tau}^{MI} = \frac{1}{M}\sum_{k=1}^M \frac{ \sum \sum_{i<j} \phi_{ij}^{(k)}}{n(n-1)/2},
\]
where
\begin{equation} \nonumber
\phi_{ij}^{(k)} = \left\{ \begin{array}{l l}
1 & \mbox{if } R_{i,(k)}^X < L_{j,(k)}^X, R_{i,(k)}^Y < L_{j,(k)}^Y \mbox{ or } R_{j,(k)}^X < L_{i,(k)}^X, R_{j,(k)}^Y < L_{i,(k)}^Y, \\
-1 & \mbox{if } R_{i,(k)}^X < L_{j,(k)}^X, L_{i,(k)}^Y > R_{j,(k)}^Y \mbox{ or } R_{j,(k)}^X < L_{i,(k)}^X,
L_{j,(k)}^Y > R_{i,(k)}^Y, \\
0 & \mbox{otherwise}.
\end{array} \right.
\end{equation}
In this paper, we set $M=100$ following \cite{Betensky:1999b}.
With three measures $\hat{\tau}^{MI}$, $\hat{\tau}^{NP}$, and $\hat{\tau}^{RP}$,
we define test statistics $Z_{MI}
=\hat{\tau}^{MI}/\sqrt{{\rm Var}(\hat{\tau}^{MI})}$,
$Z_{NP} = \hat{\tau}^{NP}/\sqrt{{\rm Var}(\hat{\tau}^{NP})}$,
$Z_{RP} = \hat{\tau}^{RP}/\sqrt{{\rm Var}(\hat{\tau}^{RP})}$.
Under the null hypothesis, the test statistics $Z_{MI}$, $Z_{NP}$, and $Z_{RP}$
follow the standard normal distribution.

As in the bivariate right-censored data,
we evaluate p-values of $\hat{\tau}^{MI}$, $\hat{\tau}^{NP}$ and $\hat{\tau}^{RP}$ with their asymptotic null distributions and the approximated
null distribution by permutation.
We summarize the bias, MSE, EP-A, and EP-P in Tables 2 and 3 for
the scenarios {\bf C1} and {\bf C2}, respectively.

From Tables 2 and 3, we can easily see that $\hat{\tau}^{NP}$
and $\hat{\tau}^{RP}$ outperform $\hat{\tau}^{MI}$ in terms
of the power of test and sizes of all three testing procedures
are well-controlled.
Thus, we focus on the comparison between $\hat{\tau}^{NP}$ and
$\hat{\tau}^{RP}$.
First, for both scenarios $\bf C1$ and $\bf C2$, $\hat{\tau}^{RP}$ has smaller bias than $\hat{\tau}^{NP}$
when the true parameter $\tau = 0$ but
has larger bias when $\tau \neq 0$. Our numerical study supports
that the estimator $\hat{\tau}^{NP}$ is an unbiased estimator of
Kendall's tau for bivariate interval-censored data.
For the ideal scenario ${\bf C2}$, two testing procedures $\hat{\tau}^{RP}$
and $\hat{\tau}^{NP}$ have quite similar powers. In this scenario,
the size of intervals is fixed and non-overlapped since all patients attend the pre-scheduled examination, and then the events are observed. Thus, $\hat{\tau}^{RP}$ and $\hat{\tau}^{NP}$ can fully utilize
the observed intervals. This characteristic is supported by
the positive increments of powers
and the decrement of biases of $\hat{\tau}^{RP}$ and $\hat{\tau}^{NP}$ in terms of
their absolute value from $\bf C1$ to $\bf C2$.
Next, for the scenario $\bf C1$, under Clayton copular,
$\hat{\tau}^{RP}$ outperforms $\hat{\tau}^{NP}$. On the other hand, under
Frank copular, the powers of $\hat{\tau}^{NP}$ are similar to
or slightly higher than those of $\hat{\tau}^{RP}$. Thus, again
as in the right-censored data, no method between
$\hat{\tau}^{RP}$ and $\hat{\tau}^{NP}$ wins the other for all
cases considered. However, we propose to use the RP statistic
in the sense that the gain of the powers from $\hat{\tau}^{NP}$
to $\hat{\tau}^{RP}$ under Clayton copula is significant, while the
power loss by using $\hat{\tau}^{RP}$ instead of $\hat{\tau}^{NP}$
is minor only under Frank copula.

\section{Data examples}

\begin{figure}[htb!]
\centering
\begin{minipage}{0.32\linewidth}
\includegraphics[width=0.9\linewidth]{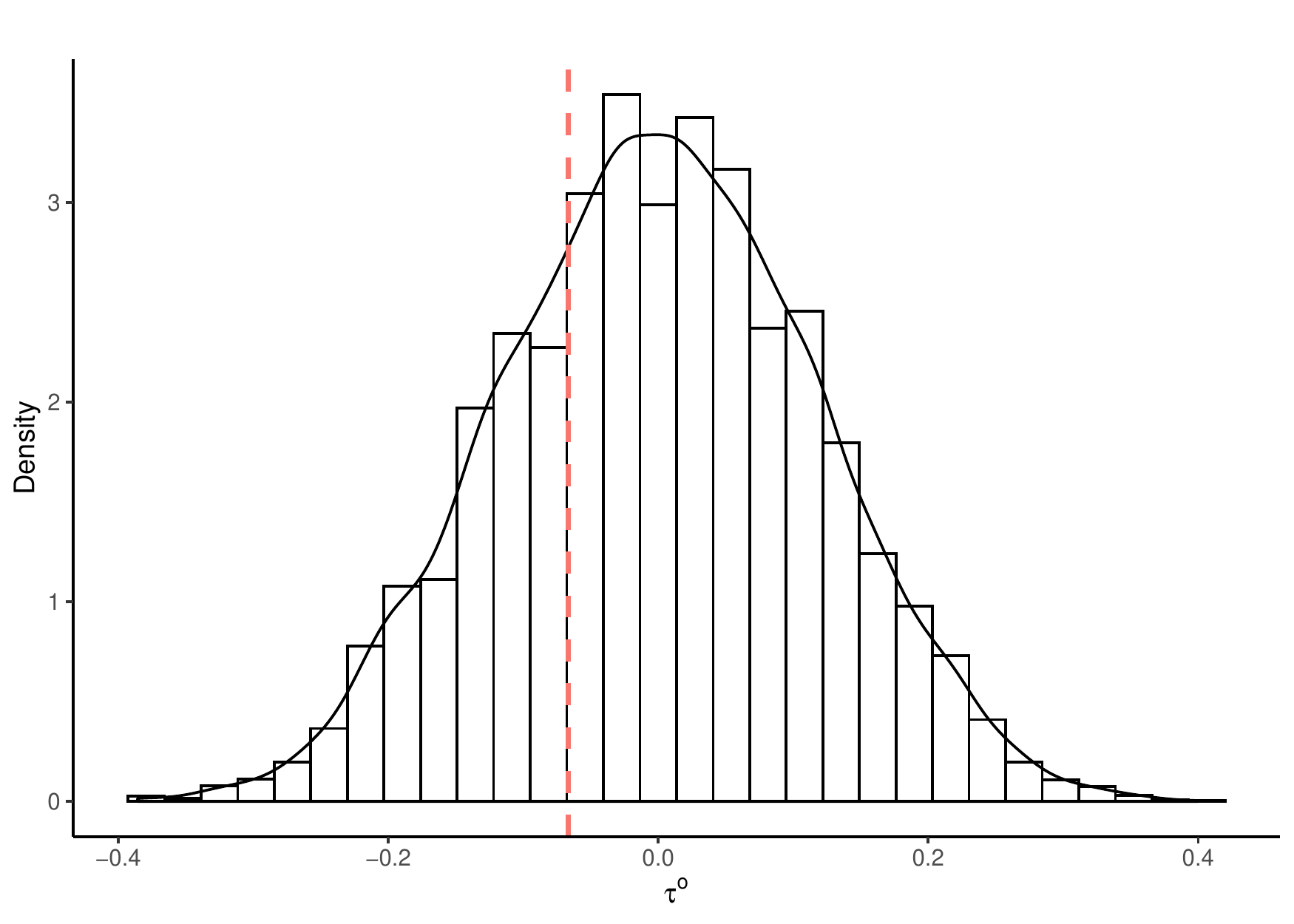} 
\centerline{(a) $\hat{\tau}^O$ }
\end{minipage} \medskip
\begin{minipage}{0.32\linewidth}
\includegraphics[width=0.9\linewidth]{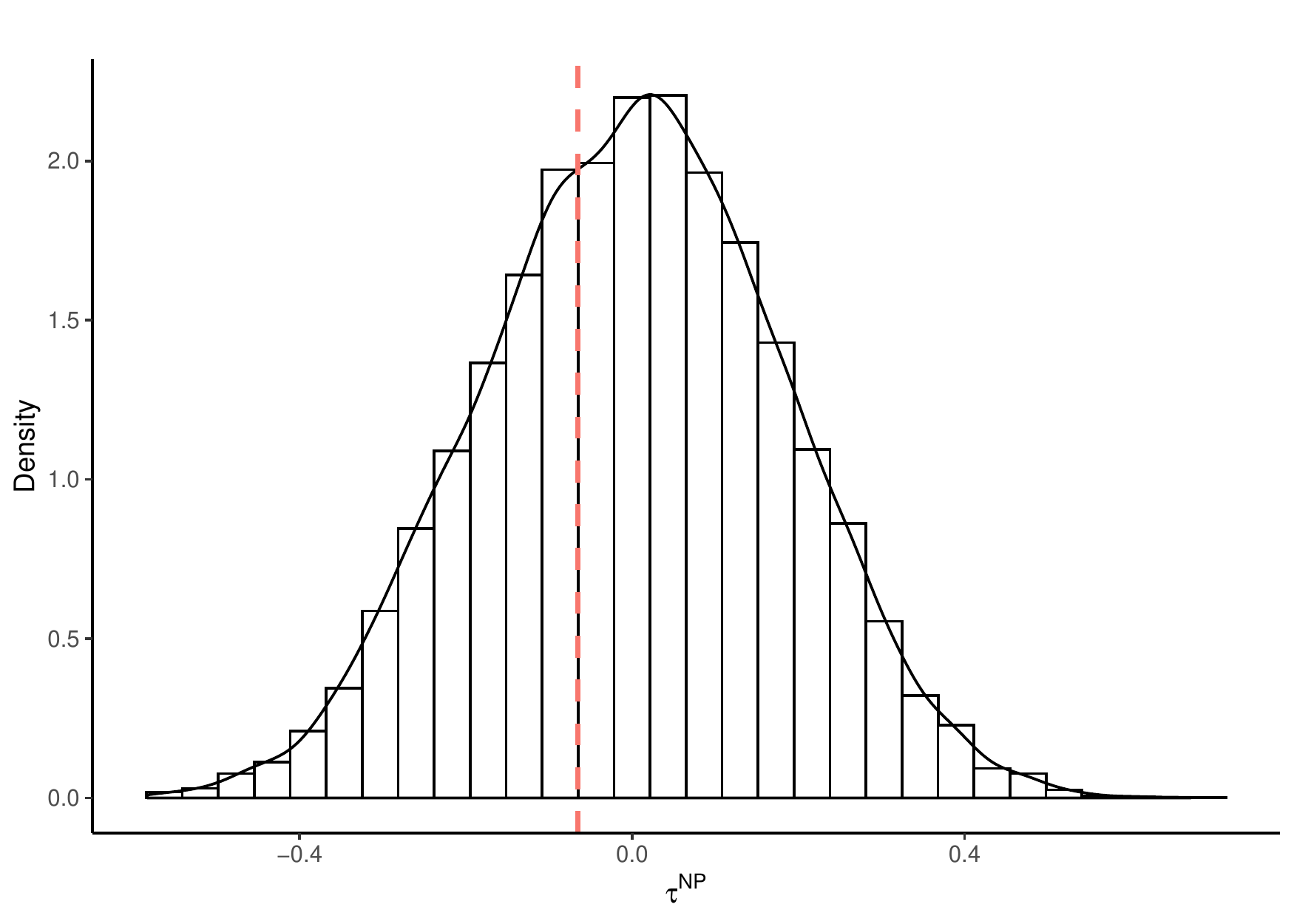}
\centerline{(b) $\hat{\tau}^{NP}$ }
\end{minipage} \medskip
\begin{minipage}{0.32\linewidth}
\includegraphics[width=0.9\linewidth]{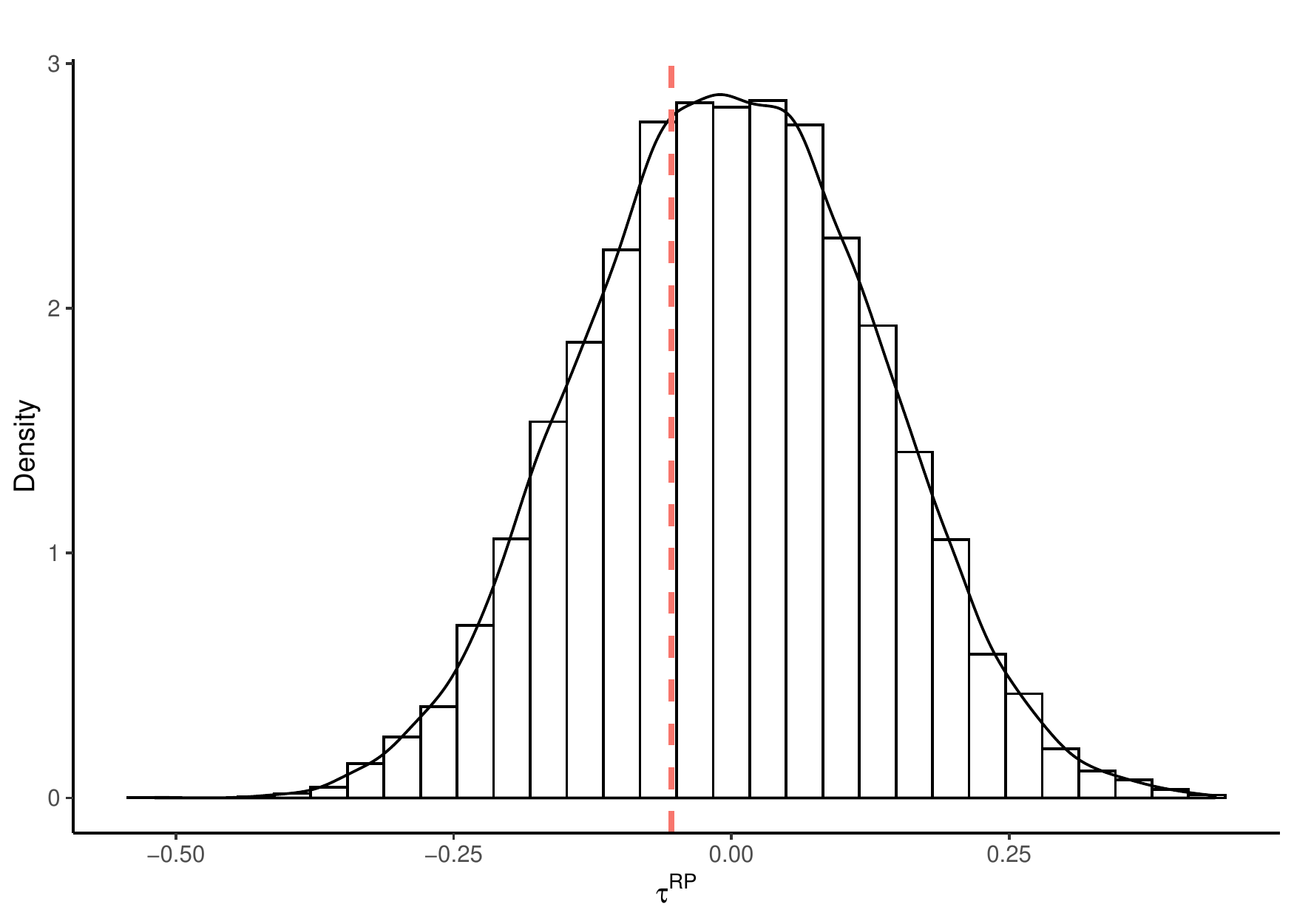} 
\centerline{(c) $\hat{\tau}^{RP}$}
\end{minipage} \medskip
\begin{minipage}{0.32\linewidth}
\includegraphics[width=0.9\linewidth]{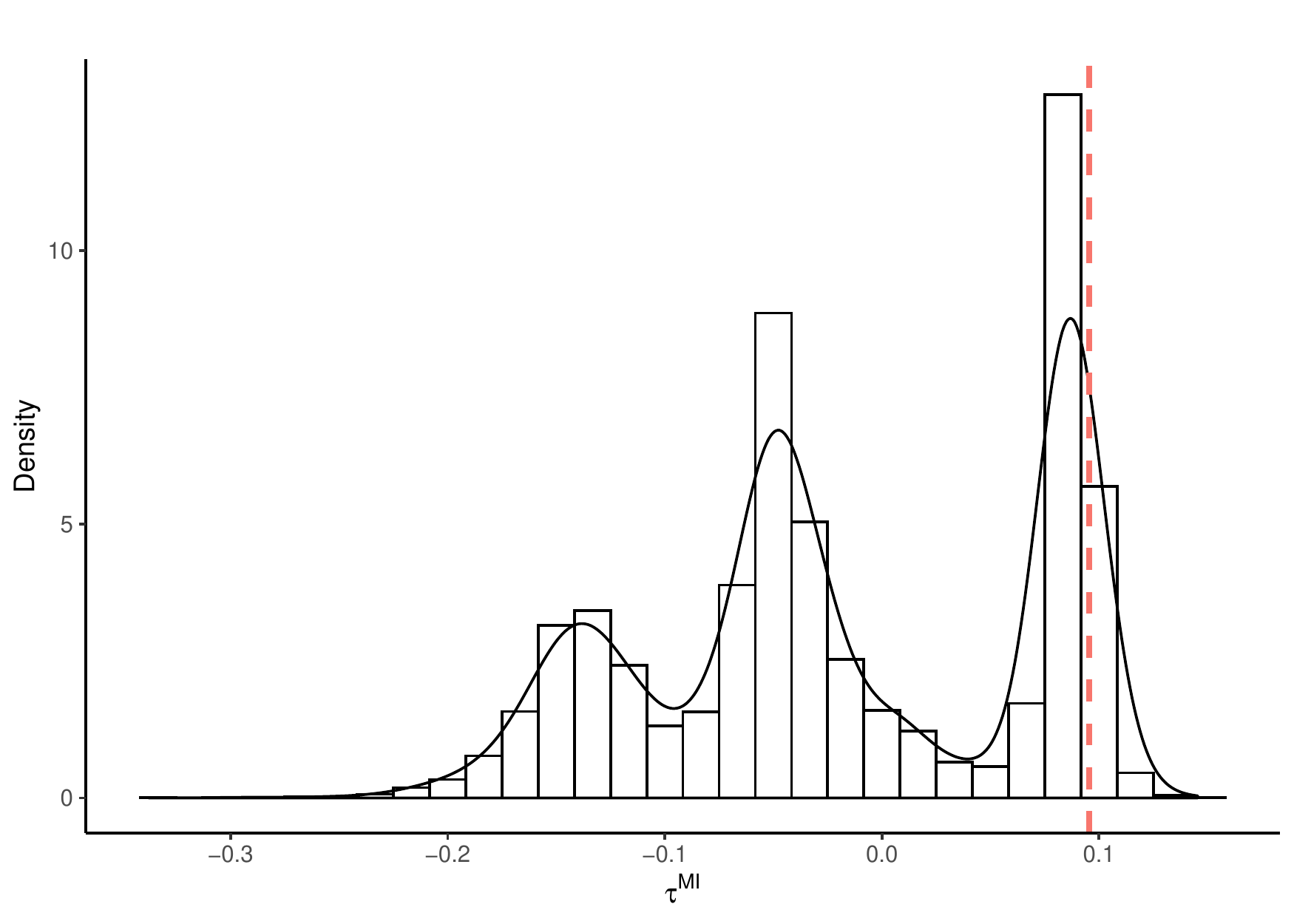}
\centerline{(d) $\hat{\tau}^{MI}$}
\end{minipage}
\begin{minipage}{0.32\linewidth}
\includegraphics[width=0.9\linewidth]{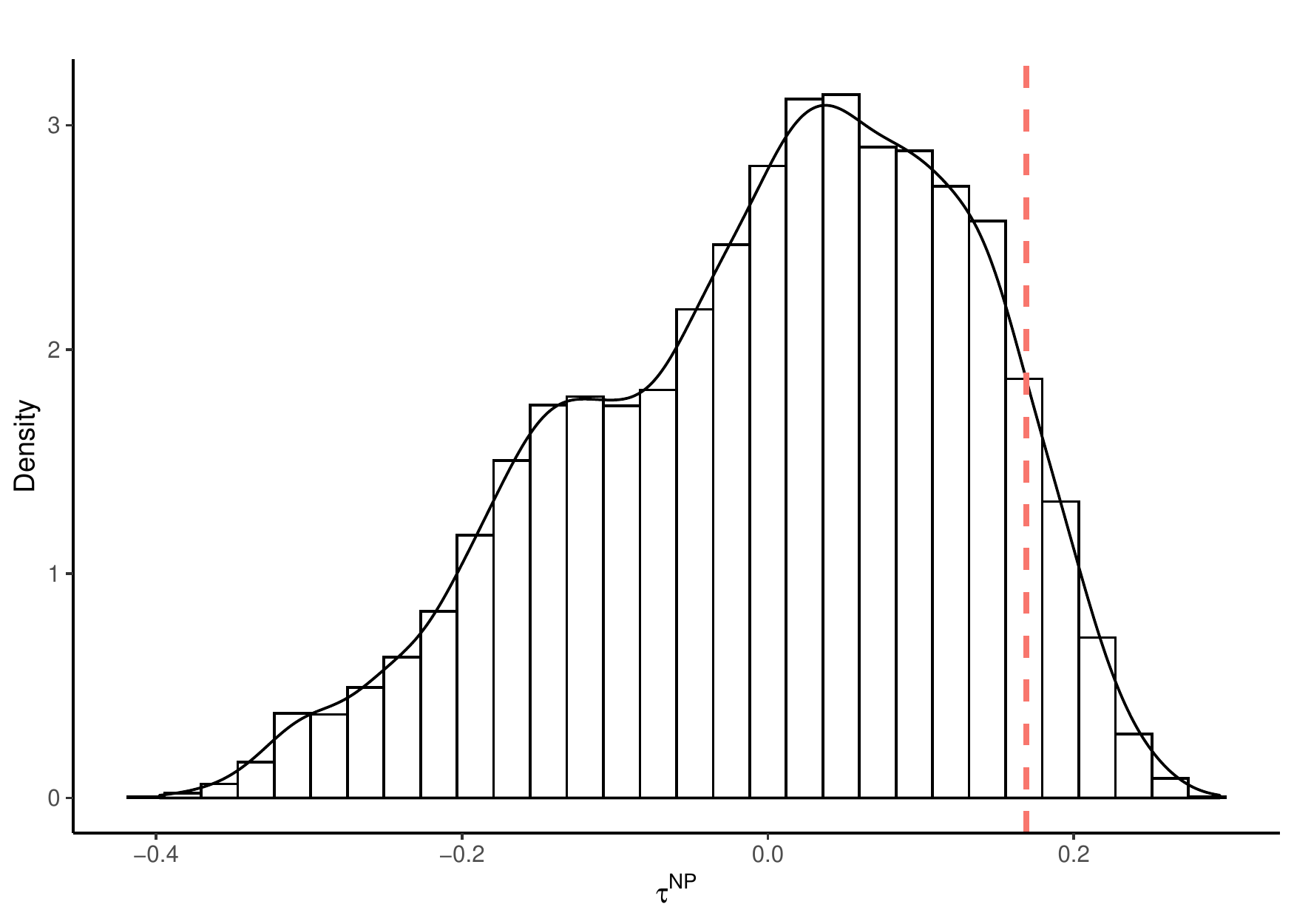}
\centerline{(e) $\hat{\tau}^{NP}$}
\end{minipage}
\begin{minipage}{0.32\linewidth}
\includegraphics[width=0.9\linewidth]{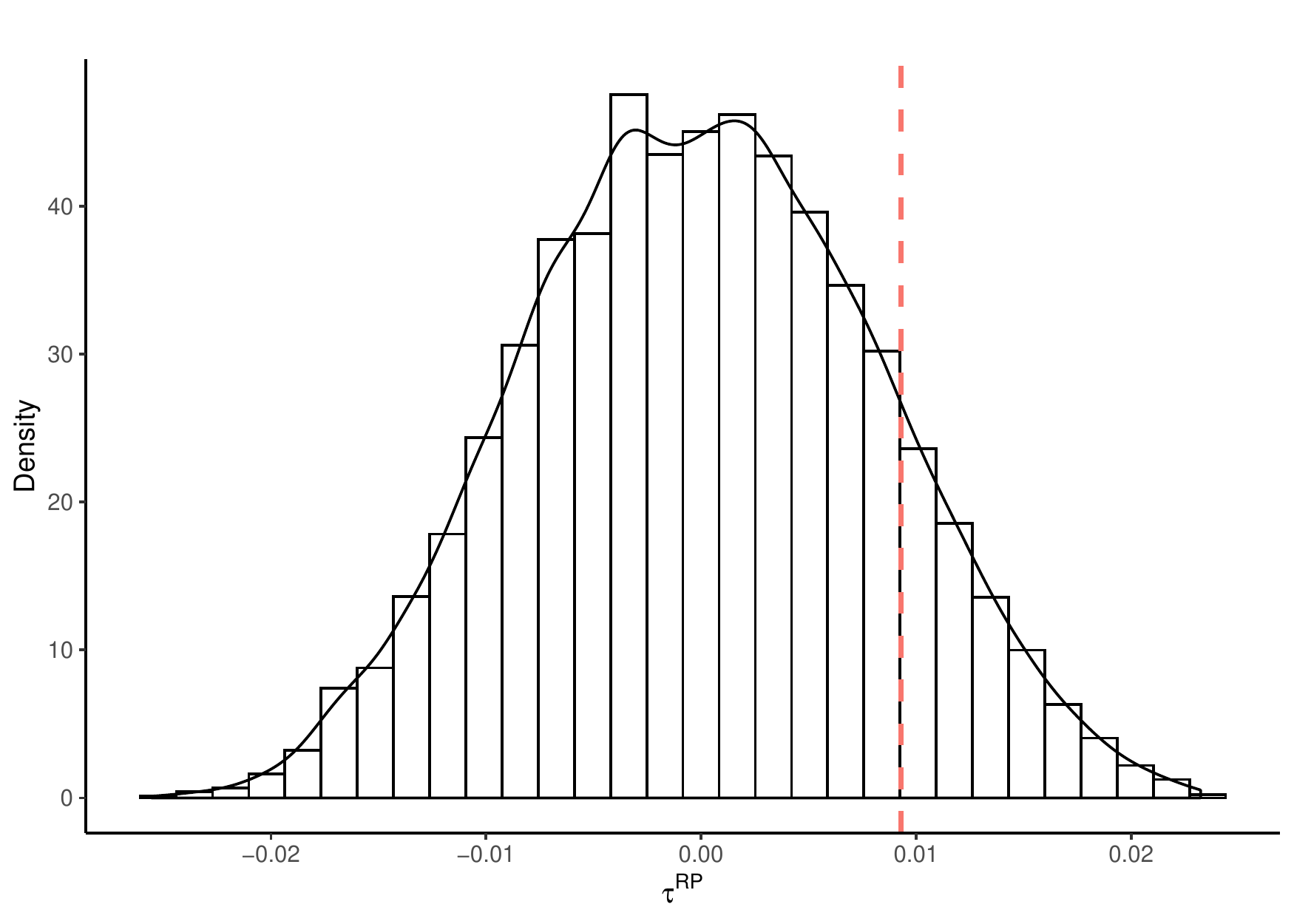}
\centerline{(f) $\hat{\tau}^{RP}$ }
\end{minipage}
\begin{minipage}{0.32\linewidth}
\includegraphics[width=0.9\linewidth]{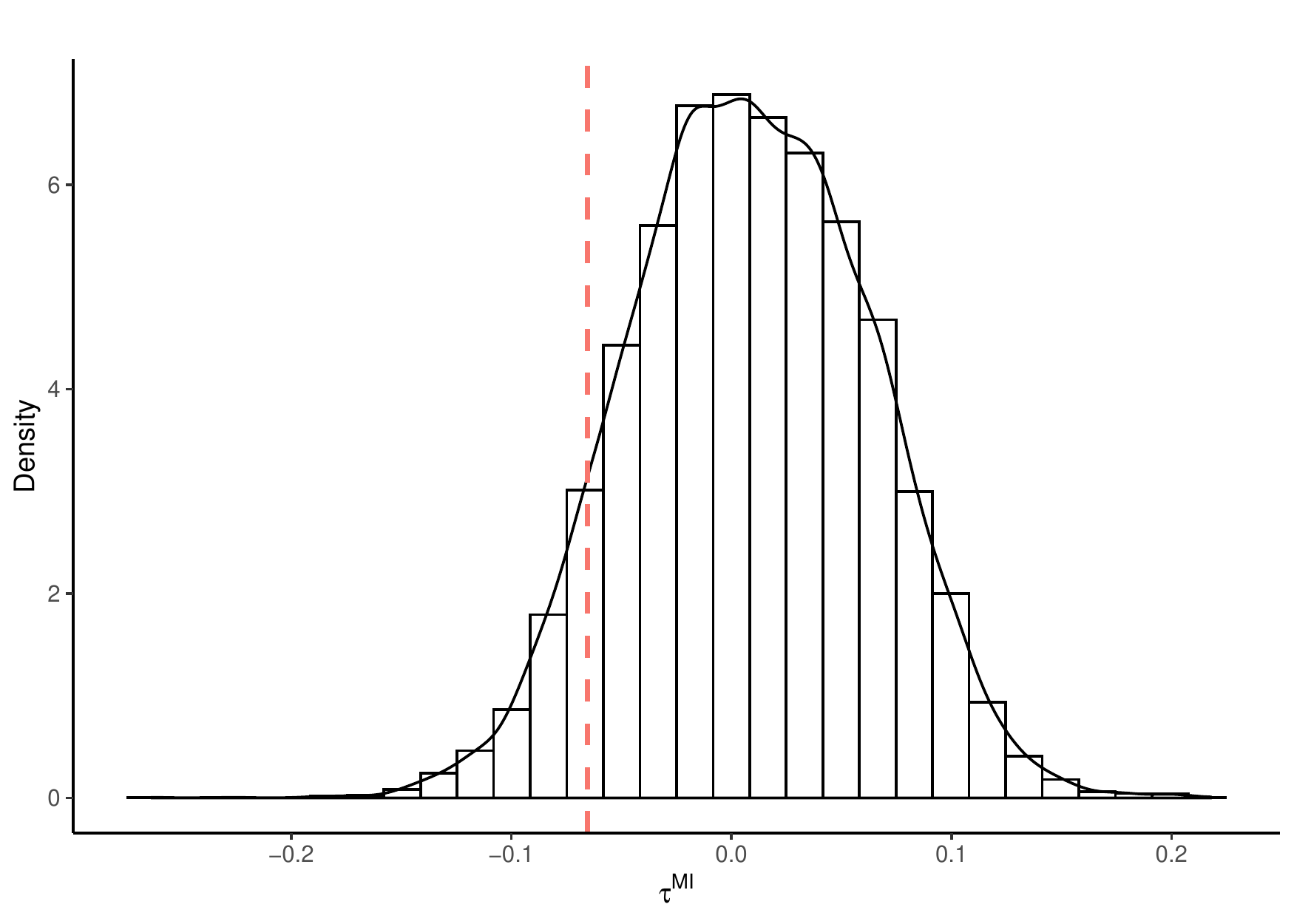}
\centerline{(g) $\hat{\tau}^{MI}$ }
\end{minipage}
\begin{minipage}{0.32\linewidth}
\includegraphics[width=0.9\linewidth]{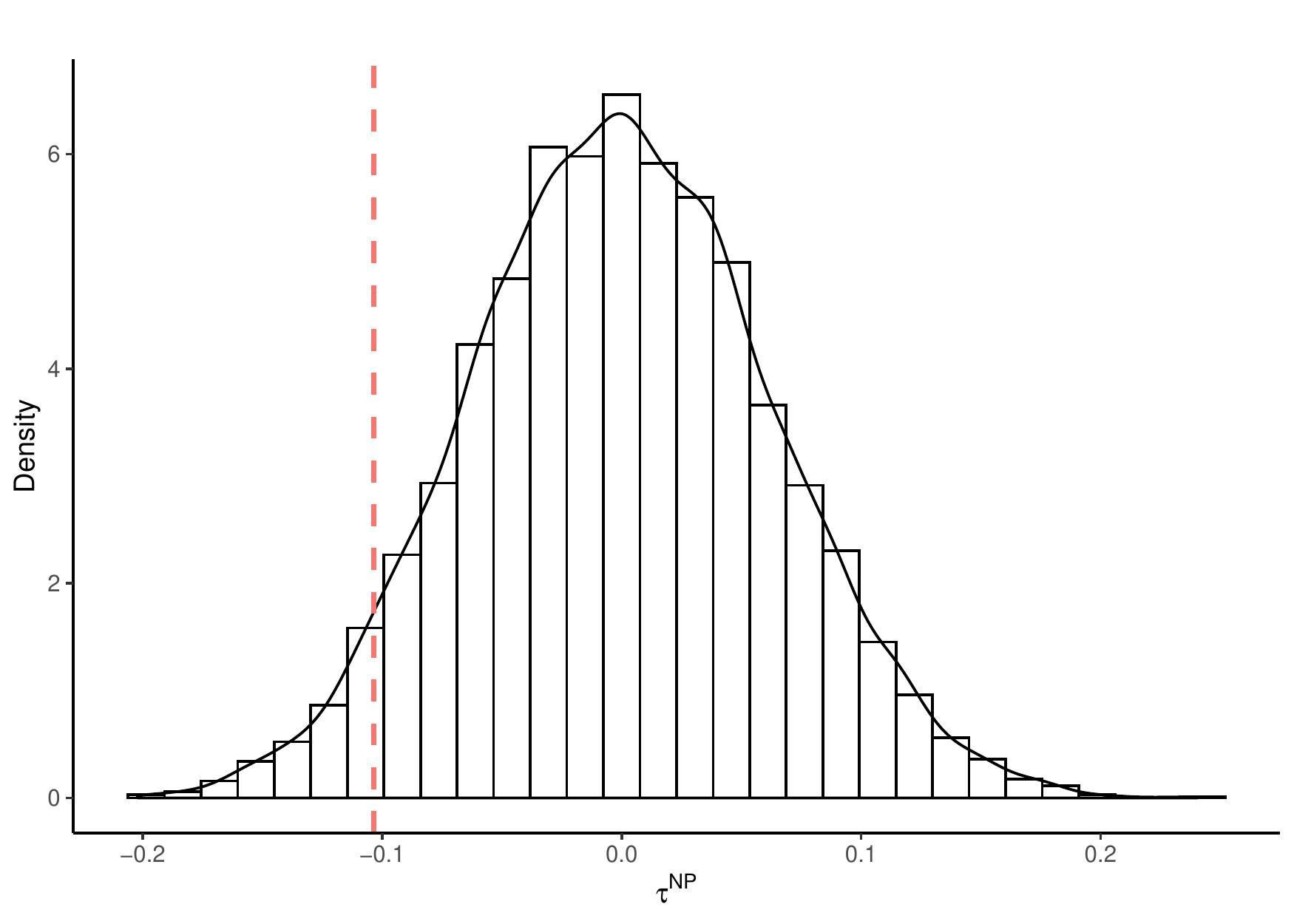}
\centerline{(h) $\hat{\tau}^{NP}$ }
\end{minipage}
\begin{minipage}{0.32\linewidth}
\includegraphics[width=0.9\linewidth]{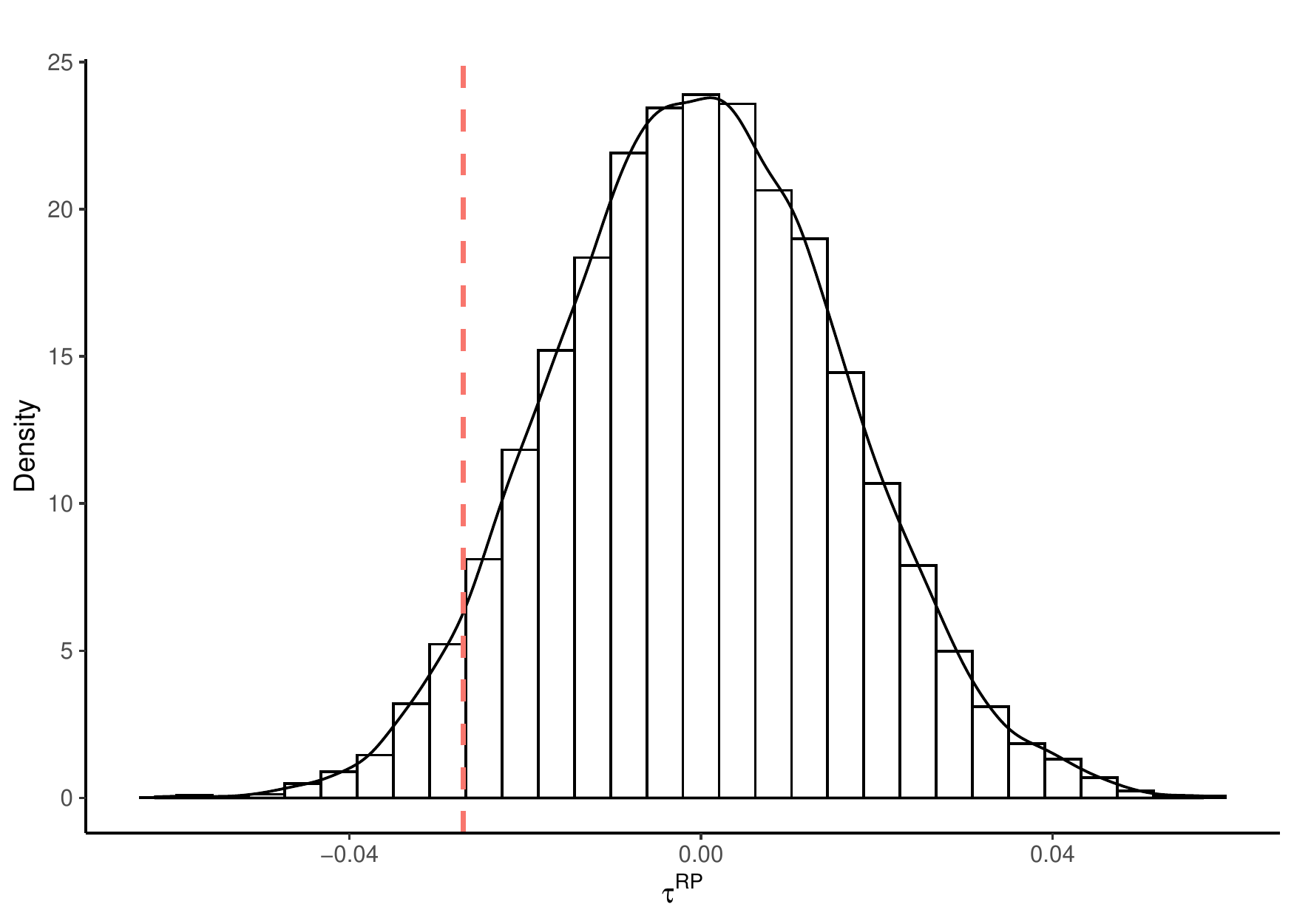}
\centerline{(i) $\hat{\tau}^{RP}$}
\end{minipage}
\caption{Density plots of null distributions of three test statistics approximated by 10,000 permutation for the leukemia remission (top), animal tumorigenicity (middle), and ACTG 181 (bottom) datasets. Left, center, and right plots represent the density plots of $\hat{\tau}^O$, $\hat{\tau}^{NP}$, $\hat{\tau}^{RP}$ for
Leukemia remission data and $\hat{\tau}^{MI}$, $\hat{\tau}^{NP}$, $\hat{\tau}^{RP}$
for Animal tumorigenicity and ACTG 181 datasets, respectively. Red vertical dashed lines represent the value of test statistics.} \label{fig:app}
\end{figure}

In this section,
to provide real-data applications, we test the independence of the
bivariate censored data with three real-data examples:
a bivariate right-censored dataset, bivariate case-1, and case-2 interval-censored datasets. In addition, to check
reliability of the proposed procedure,
we consider $\hat{\tau}^O$, $\hat{\tau}^{NP}$, and $\hat{\tau}^{RP}$ for the bivariate right-censored
data and $\hat{\tau}^{MI}$, $\hat{\tau}^{NP}$, and $\hat{\tau}^{RP}$ for the bivariate
interval-censored data as considered in the numerical study.
The variances of the null distributions of the test statistics
and the p-values evaluated by permutation are
estimated by 10,000 permutations, and the RP statistic $\hat{\tau}^{RP}$ is
obtained with 5,000 MCMC samples.
All the results of testing independence are summarized in Table \ref{tab:app},
and the null distributions approximated by permutation are plotted
in Figure \ref{fig:app}. Figure \ref{fig:app} shows that
the null distribution of the RP statistic is well-approximated by
the permutation procedure.

\subsection{Leukemia remission data (right-censored data)}

For the bivariate right-censored data,
we consider the leukemia remission data set, which is
a well-known bivariate right-censored
data analyzed in \cite{Oakes:1982}.
The leukemia remission dataset consists of 21 paired lengths
of remission maintenance with 6-MP (6-Mercaptopurine) and placebo
from Phase II analysis in \cite{Freireich:1963}.
For the purpose of the fair comparison and taking into account the existence of ties,
we use the length of remission maintenance directly rather
than their ranks in \cite{Oakes:1982}, where the ties are broken by randomization and
an assumption that the death at a given time always proceeds censoring at the same time.

In the leukemia remission data,
the estimate $\hat{\tau}^{O}$ of Kendall's tau by \cite{Oakes:1982} is $-0.0667$, and its estimated variance by permutation is $0.0136$.
The p-values evaluated by the asymptotic null distribution (p-Asym) and
permutation (p-Perm) of $\hat{\tau}^{O}$ are $0.5680$ and $0.5748$, respectively.
For $\hat{\tau}^{NP}$, the estimate of Kendall's tau by \cite{Kim:2015} is $-0.0652$, and its variance is estimated as $0.0322$.
The p-Asym and p-Perm of $\hat{\tau}^{NP}$ are $0.7163$ and $0.7200$, respectively.
For $\hat{\tau}^{RP}$, the estimate of the proposed RP statistic is $-0.0540$, and its variance estimate is $0.0170$.
The p-Asym and p-Perm of $\hat{\tau}^{RP}$ are $0.6787$ and $0.6998$, respectively.
The estimate of RP statistic is relatively smaller than the others $\hat{\tau}^{O}$
and $\hat{\tau}^{NP}$ in terms of the absolute value.
All three testing procedures consistently indicate that
there is no evidence against $H_0$, which is the same conclusion
of testing independence with the ranks of length
of remission in \cite{Oakes:1982}.

\subsection{Animal tumorigenicity data (case-1 interval-censored data)}

For the case-1 interval-censored data,
we consider the animal tumorigenicity experiment dataset from National Toxicology
Program (NTP) analyzed in \cite{Dunson:2002}.
The experiment is designed to identify the effects of a chloroprene
for carcinogenicity with 200 mice, which are separated into four groups,
each of which consists of 50 mice and is exposed
to the chloroprene of concentration 0, 12.8, 32, or 80 ppm
for up to 2 years.
In the experiment, all mice are either died by the natural cause
or sacrificed at the end of 2 years. After death, the occurrence
of tumors in various sites was determined through a pathologic examination.

In \cite{Dunson:2002}, the dataset provides
the age at death and the number of mice at given age at death
having no tumors, adrenal site only, lung site only, both sites.
In this example, for the purpose of illustration of
the inter-dependency,
we test the independence of the occurrence times of tumors
at the adrenal and the lung sites with the control group
(chloroprene of concentration 0).

In the animal tumorigenicity data,
the estimate $\hat{\tau}^{MI}$ of Kendall's tau by \cite{Betensky:1999b} is $0.0955$, and its estimated variance by permutation is $0.0080$.
The p-Asym and p-Perm of $\hat{\tau}^{MI}$ are $0.2851$ and $0.2046$, respectively.
For $\hat{\tau}^{NP}$, the estimate of the Kendall's tau by \cite{Kim:2015} is $0.1689$ and its variance estimate is $0.0163$.
The p-Asym and p-Perm of $\hat{\tau}^{NP}$ are $0.1853$ and $0.1915$, respectively.
For $\hat{\tau}^{RP}$, the estimate of the proposed RP statistic is $0.0093$, and its variance estimate is $0.0001$.
The p-Asym and p-Perm of $\hat{\tau}^{RP}$ are $0.2519$ and $0.2641$, respectively.
As observed in the numerical study,
the p-Asym of $\hat{\tau}^{MI}$ is greater than the p-Perm of $\hat{\tau}^{MI}$
while the p-Asym and the p-Perm of the other two methods are similar.
From the results of three testing independence procedures,
there is no evidence that the occurrence times of cancer at
the adrenal and the lung sites are correlated.
Remark that the results of $\hat{\tau}^{MI}$ based on the null distribution
approximated by permutation is consistent with others, but the permutation
procedure poorly estimates the null distribution of $\hat{\tau}^{MI}$ as
shown in Figure \ref{fig:app}.

\subsection{ACTG 181 data (case-2 interval-censored data)}

For the case-2 interval-censored data,
we consider the AIDS Clinical Trials Group protocol (ACTG 181) data analyzed by \cite{Betensky:1999}.
This study aims at identifying an association between the first time
of shedding of cytomegalovirus (CMV) and colonization of mycobacterium avium complex (MAC).
Since the investigation is conducted in the laboratory and many patients missed several pre-scheduled clinical visits,
the observations of the first time of the event are interval-censored.
In \cite{Betensky:1999}, the association is estimated with 204 subjects which are tested for CMV shedding and
MAC colonization at least once during the trial and
did not have a prior CMV or MAC diagnosis.
As in the analysis by \cite{Betensky:1999}, we estimate $\hat{\tau}^{MI}$ with 100 imputed datasets.

In the ACTG 181 data,
the estimate $\hat{\tau}^{MI}$ is $-0.0654$ and its estimated variance by permutation is $0.0029$.
The p-Asym and p-Perm of $\hat{\tau}^{MI}$ are $0.2260$ and $0.1706$, respectively.
For $\hat{\tau}^{NP}$, the estimate of Kendall's tau is $-0.1035$, and its variance estimate is $0.0038$.
The p-Asym and p-Perm of $\hat{\tau}^{NP}$ are $0.0957$ and $0.0960$, respectively.
For $\hat{\tau}^{RP}$, the estimate of the proposed RP statistic is $-0.0270$, and its variance estimate is $0.0003$.
The p-Asym and p-Perm of $\hat{\tau}^{RP}$ are $0.0962$ and $0.0954$, respectively.
Although the difference between the estimates $\hat{\tau}^{NP}$ and $\hat{\tau}^{RP}$ is relatively large (0.0765), the p-Asyms and p-Perms
of $\hat{\tau}^{RP}$ and $\hat{\tau}^{NP}$ have similar values around 0.0960, which indicates that $\hat{\tau}^{RP}$ and $\hat{\tau}^{NP}$ lead the same conclusion
that there is a significant correlation between the times of CMV shedding and
MAC colonization when we consider the significance level $\alpha$ as 0.10.
As stated in \cite{Betensky:1999b},
the results of three testing procedures
indicate either a weak negative association between
the times of CMV shedding and MAC colonization or the lack of sufficient
power to detect the association.

\section{Conclusion}

In this paper, we propose a rank-based procedure to test the independence for bivariate incomplete data, which 
can be applicable to any type of incomplete data. The incomplete data defines a restricted permutation space
of all possible ranks of the observations. We propose the average of Kendall's tau over the ranks in the restricted permutation space as our test statistic. To evaluate the statistic and its reference distribution, we develop the MCMC procedure that obtains uniform samples on the restricted permutation space,
which is applied to approximating the null distribution of the averaged Kendall's tau.

Our proposal, the RP statistic, in this paper is generic and flexible enough in two ways. First, it is applicable to any type of incomplete data.
Here, we focus on testing the independence of various types of bivariate censored data, which has been studied by different groups of researchers in the literature. Again, for the censored data, no
assumptions are made for the underlying distribution and censoring mechanism. The RP statistic
performs as good as other state-of-the-art methods for censoring mechanisms considered. Second, in this paper, we use the average of the uniform samples from the
restricted rank space for comparison with the score-based procedure. However, there are many ways to make the summary of the uniform samples. Some
examples would be the quantiles, median, or L-statistics, and they may perform better than the proposed
RP statistic. This is left for future work.

\begin{table}[htb!]
\centering
\caption{Biases, mean squared error (MSE), 
empirical power evaluated by asymptotic null distribution (EP-A), and
empirical power evaluated by permutation (EP-P) for bivariate right-censored data.} \medskip
\scalebox{0.95}{
\begin{tabular}{cccccccccccc} \hline
\multirow{2}{*}{$c_R$} &\multirow{2}{*}{$n$}  & \multirow{2}{*}{$\tau$} &\multirow{2}{*}{Est.} &
\multicolumn{4}{c}{Clayton} & \multicolumn{4}{c}{Frank} \\ \cline{5-12}
&  &  &  & Bias & MSE & EP-A & EP-P & Bias & MSE & EP-A & EP-B \\ \hline
\multirow{18}{*}{$9$} & \multirow{9}{*}{$50$} & \multirow{3}{*}{$0$} & $\hat{\tau}^{O}$ & 0.0037 & 0.0016 & 0.036 & 0.034 & 0.0038 & 0.0016 & 0.036 & 0.034 \\
 &  &  & $\hat{\tau}^{NP}$ & 0.0093 & 0.0142 & 0.042 & 0.040 & 0.0096 & 0.0142 & 0.042 & 0.040 \\
 &  &  & $\tilde{\tau}^{RP}$ & 0.0042 & 0.0024 & 0.044 & 0.048 & 0.0045 & 0.0024 & 0.044 & 0.048 \\ \cline{3-12}
 &  & \multirow{3}{*}{$1/5$} & $\hat{\tau}^{O}$ & -0.1078 & 0.0145 & 0.612 & 0.616 & -0.1471 & 0.0239 & 0.276 & 0.286 \\
 &  &  & $\hat{\tau}^{NP}$ & 0.0299 & 0.0192 & 0.504 & 0.506 & -0.0385 & 0.0188 & 0.266 & 0.262 \\
 &  &  & $\tilde{\tau}^{RP}$ & -0.0978 & 0.0132 & 0.556 & 0.552 & -0.1389 & 0.0223 & 0.276 & 0.276 \\ \cline{3-12}
 &  & \multirow{3}{*}{$1/3$} & $\hat{\tau}^{O}$ & -0.1859 & 0.0380 & 0.890 & 0.896 & -0.2407 & 0.0604 & 0.606 & 0.608 \\
 &  &  & $\hat{\tau}^{NP}$ & 0.0393 & 0.0173 & 0.842 & 0.842 & -0.0568 & 0.0189 & 0.644 & 0.648 \\
 &  &  & $\tilde{\tau}^{RP}$ & -0.1687 & 0.0324 & 0.874 & 0.876 & -0.2263 & 0.0544 & 0.590 & 0.588 \\ \cline{2-12}
 & \multirow{9}{*}{$100$} & \multirow{3}{*}{$0$} & $\hat{\tau}^{O}$ & -0.0018 & 0.0008 & 0.044 & 0.046 & -0.0018 & 0.0008 & 0.044 & 0.046 \\
 &  &  & $\hat{\tau}^{NP}$ & -0.0043 & 0.0069 & 0.040 & 0.042 & -0.0046 & 0.0069 & 0.040 & 0.042 \\
 &  &  & $\tilde{\tau}^{RP}$ & -0.0017 & 0.0011 & 0.036 & 0.038 & -0.0018 & 0.0011 & 0.036 & 0.038 \\ \cline{3-12}
 &  & \multirow{3}{*}{$1/5$} & $\hat{\tau}^{O}$ & -0.1146 & 0.0145 & 0.798 & 0.804 & -0.1415 & 0.0210 & 0.518 & 0.520 \\
 &  &  & $\hat{\tau}^{NP}$ & 0.0205 & 0.0091 & 0.716 & 0.714 & -0.0244 & 0.0077 & 0.546 & 0.544 \\
 &  &  & $\tilde{\tau}^{RP}$ & -0.1044 & 0.0126 & 0.762 & 0.766 & -0.1321 & 0.0189 & 0.480 & 0.478 \\ \cline{3-12}
 &  & \multirow{3}{*}{$1/3$} & $\hat{\tau}^{O}$ & -0.1916 & 0.0385 & 0.988 & 0.984 & -0.2356 & 0.0566 & 0.934 & 0.934 \\
 &  &  & $\hat{\tau}^{NP}$ & 0.0344 & 0.0095 & 0.972 & 0.972 & -0.0449 & 0.0086 & 0.936 & 0.932 \\
 &  &  & $\tilde{\tau}^{RP}$ & -0.1742 & 0.0324 & 0.984 & 0.984 & -0.2201 & 0.0499 & 0.908 & 0.908 \\ \hline
\multirow{18}{*}{$15$} & \multirow{9}{*}{$50$} & \multirow{3}{*}{$0$} & $\hat{\tau}^{O}$ & 0.0036 & 0.0029 & 0.036 & 0.040 & 0.0036 & 0.0029 & 0.036 & 0.040 \\
 &  &  & $\hat{\tau}^{NP}$ & 0.0057 & 0.0132 & 0.038 & 0.034 & 0.0056 & 0.0132 & 0.038 & 0.034 \\
 &  &  & $\tilde{\tau}^{RP}$ & 0.0037 & 0.0041 & 0.038 & 0.046 & 0.0037 & 0.0041 & 0.038 & 0.046 \\ \cline{3-12}
 &  & \multirow{3}{*}{$1/5$} & $\hat{\tau}^{O}$ & -0.0798 & 0.0105 & 0.582 & 0.576 & -0.1160 & 0.0169 & 0.356 & 0.368 \\
 &  &  & $\hat{\tau}^{NP}$ & 0.0221 & 0.0149 & 0.498 & 0.494 & -0.0184 & 0.0141 & 0.348 & 0.348 \\
 &  &  & $\tilde{\tau}^{RP}$ & -0.0694 & 0.0098 & 0.526 & 0.520 & -0.1040 & 0.0155 & 0.358 & 0.348 \\ \cline{3-12}
 &  & \multirow{3}{*}{$1/3$} & $\hat{\tau}^{O}$ & -0.1388 & 0.0239 & 0.934 & 0.936 & -0.1889 & 0.0392 & 0.734 & 0.750 \\
 &  &  & $\hat{\tau}^{NP}$ & 0.0315 & 0.0135 & 0.886 & 0.884 & -0.0270 & 0.0131 & 0.754 & 0.758 \\
 &  &  & $\tilde{\tau}^{RP}$ & -0.1201 & 0.0197 & 0.916 & 0.914 & -0.1679 & 0.0327 & 0.728 & 0.734 \\ \cline{2-12}
 & \multirow{9}{*}{$100$} & \multirow{3}{*}{$0$} & $\hat{\tau}^{O}$ & -0.0018 & 0.0015 & 0.048 & 0.048 & -0.0020 & 0.0015 & 0.048 & 0.048 \\
 &  &  & $\hat{\tau}^{NP}$ & -0.0032 & 0.0066 & 0.050 & 0.052 & -0.0035 & 0.0067 & 0.050 & 0.052 \\
 &  &  & $\tilde{\tau}^{RP}$ & -0.0014 & 0.0020 & 0.050 & 0.052 & -0.0016 & 0.0021 & 0.052 & 0.054 \\ \cline{3-12}
 &  & \multirow{3}{*}{$1/5$} & $\hat{\tau}^{O}$ & -0.0873 & 0.0097 & 0.810 & 0.818 & -0.1084 & 0.0133 & 0.658 & 0.662 \\
 &  &  & $\hat{\tau}^{NP}$ & 0.0140 & 0.0079 & 0.728 & 0.722 & -0.0044 & 0.0061 & 0.700 & 0.690 \\
 &  &  & $\tilde{\tau}^{RP}$ & -0.0754 & 0.0084 & 0.768 & 0.766 & -0.0957 & 0.0113 & 0.634 & 0.638 \\ \cline{3-12}
 &  & \multirow{3}{*}{$1/3$} & $\hat{\tau}^{O}$ & -0.1461 & 0.0237 & 0.994 & 0.994 & -0.1839 & 0.0354 & 0.982 & 0.982 \\
 &  &  & $\hat{\tau}^{NP}$ & 0.0243 & 0.0074 & 0.992 & 0.992 & -0.0167 & 0.0055 & 0.984 & 0.984 \\
 &  &  & $\tilde{\tau}^{RP}$ & -0.1262 & 0.0187 & 0.990 & 0.990 & -0.1628 & 0.0286 & 0.966 & 0.972 \\ \hline
\end{tabular}
}
\end{table}

\begin{table}[htb!]
\centering
\caption{Biases, mean squared error (MSE), 
empirical power evaluated by asymptotic null distribution (EP-A), and
empirical power evaluated by permutation (EP-P) for bivariate interval-censored data under the given scenario {\bf (C1)}.} \medskip
\scalebox{0.95}{
\begin{tabular}{cccccccccccc} \hline
\multirow{2}{*}{$c_R$} &\multirow{2}{*}{$n$}  & \multirow{2}{*}{$\tau$} &\multirow{2}{*}{Est.} &
\multicolumn{4}{c}{Clayton} & \multicolumn{4}{c}{Frank} \\ \cline{5-12}
&  &  &  & Bias & MSE & EP-A & EP-P & Bias & MSE & EP-A & EP-B \\ \hline
\multirow{18}{*}{$6$} & \multirow{9}{*}{$50$} & \multirow{3}{*}{$0$} & $\hat{\tau}^{MI}$ & 0.0379 & 0.0362 & 0.042 & 0.054 & 0.0339 & 0.0342 & 0.064 & 0.058 \\
 &  &  & $\hat{\tau}^{NP}$ & 0.0043 & 0.0197 & 0.054 & 0.058 & 0.0008 & 0.0193 & 0.052 & 0.054 \\
 &  &  & $\hat{\tau}^{RP}$ & 0.0007 & 0.0014 & 0.058 & 0.062 & 0.0014 & 0.0013 & 0.050 & 0.050 \\ \cline{3-12}
 &  & \multirow{3}{*}{$1/5$} & $\hat{\tau}^{MI}$ & 0.0163 & 0.0315 & 0.250 & 0.218 & -0.0024 & 0.0285 & 0.212 & 0.180 \\
 &  &  & $\hat{\tau}^{NP}$ & 0.0286 & 0.0213 & 0.408 & 0.418 & -0.0212 & 0.0203 & 0.288 & 0.288 \\
 &  &  & $\hat{\tau}^{RP}$ & -0.1342 & 0.0198 & 0.446 & 0.454 & -0.1547 & 0.0255 & 0.280 & 0.284 \\ \cline{3-12}
 &  & \multirow{3}{*}{$1/3$} & $\hat{\tau}^{MI}$ & 0.0294 & 0.0235 & 0.564 & 0.498 & -0.0112 & 0.0226 & 0.462 & 0.394 \\
 &  &  & $\hat{\tau}^{NP}$ & 0.0715 & 0.0218 & 0.844 & 0.848 & -0.0166 & 0.0154 & 0.670 & 0.676 \\
 &  &  & $\hat{\tau}^{RP}$ & -0.2185 & 0.0500 & 0.890 & 0.900 & -0.2534 & 0.0657 & 0.616 & 0.630 \\ \cline{2-12}
 & \multirow{9}{*}{$100$} & \multirow{3}{*}{$0$} & $\hat{\tau}^{MI}$ & 0.0385 & 0.0230 & 0.052 & 0.052 & 0.0369 & 0.0226 & 0.044 & 0.042 \\
 &  &  & $\hat{\tau}^{NP}$ & -0.0005 & 0.0086 & 0.050 & 0.052 & 0.0028 & 0.0083 & 0.042 & 0.044 \\
 &  &  & $\hat{\tau}^{RP}$ & -0.0009 & 0.0006 & 0.052 & 0.056 & 0.0008 & 0.0006 & 0.054 & 0.054 \\ \cline{3-12}
 &  & \multirow{3}{*}{$1/5$} & $\hat{\tau}^{MI}$ & 0.0359 & 0.0223 & 0.396 & 0.304 & 0.0229 & 0.0189 & 0.358 & 0.242 \\
 &  &  & $\hat{\tau}^{NP}$ & 0.0396 & 0.0110 & 0.700 & 0.704 & -0.0139 & 0.0093 & 0.516 & 0.522 \\
 &  &  & $\hat{\tau}^{RP}$ & -0.1308 & 0.0180 & 0.770 & 0.774 & -0.1523 & 0.0239 & 0.478 & 0.472 \\ \cline{3-12}
 &  & \multirow{3}{*}{$1/3$} & $\hat{\tau}^{MI}$ & 0.0422 & 0.0196 & 0.758 & 0.680 & 0.0170 & 0.0152 & 0.716 & 0.620 \\
 &  &  & $\hat{\tau}^{NP}$ & 0.0634 & 0.0118 & 0.988 & 0.988 & -0.0147 & 0.0085 & 0.916 & 0.916 \\
 &  &  & $\hat{\tau}^{RP}$ & -0.2212 & 0.0499 & 0.994 & 0.994 & -0.2514 & 0.0640 & 0.908 & 0.912 \\ \hline
\multirow{18}{*}{$12$} & \multirow{9}{*}{$50$} & \multirow{3}{*}{$0$} & $\hat{\tau}^{MI}$ & 0.0077 & 0.0282 & 0.048 & 0.048 & 0.0022 & 0.0298 & 0.052 & 0.048 \\
 &  &  & $\hat{\tau}^{NP}$ & 0.0018 & 0.0171 & 0.050 & 0.052 & -0.0071 & 0.0190 & 0.058 & 0.064 \\
 &  &  & $\hat{\tau}^{RP}$ & 0.0010 & 0.0012 & 0.044 & 0.042 & -0.0016 & 0.0013 & 0.072 & 0.070 \\ \cline{3-12}
 &  & \multirow{3}{*}{$1/5$} & $\hat{\tau}^{MI}$ & 0.0170 & 0.0304 & 0.274 & 0.262 & 0.0086 & 0.0256 & 0.238 & 0.230 \\
 &  &  & $\hat{\tau}^{NP}$ & 0.0225 & 0.0186 & 0.368 & 0.372 & -0.0061 & 0.0164 & 0.292 & 0.296 \\
 &  &  & $\hat{\tau}^{RP}$ & -0.1379 & 0.0205 & 0.454 & 0.456 & -0.1509 & 0.0240 & 0.288 & 0.284 \\ \cline{3-12}
 &  & \multirow{3}{*}{$1/3$} & $\hat{\tau}^{MI}$ & 0.0049 & 0.0244 & 0.532 & 0.510 & 0.0111 & 0.0238 & 0.548 & 0.544 \\
 &  &  & $\hat{\tau}^{NP}$ & 0.0206 & 0.0149 & 0.768 & 0.778 & -0.0175 & 0.0152 & 0.668 & 0.666 \\
 &  &  & $\hat{\tau}^{RP}$ & -0.2337 & 0.0563 & 0.842 & 0.836 & -0.2510 & 0.0645 & 0.660 & 0.668 \\ \cline{2-12}
 & \multirow{9}{*}{$100$} & \multirow{3}{*}{$0$} & $\hat{\tau}^{MI}$ & 0.0016 & 0.0150 & 0.046 & 0.040 & 0.0102 & 0.0159 & 0.070 & 0.060 \\
 &  &  & $\hat{\tau}^{NP}$ & -0.0023 & 0.0085 & 0.044 & 0.042 & 0.0028 & 0.0087 & 0.048 & 0.050 \\
 &  &  & $\hat{\tau}^{RP}$ & 0.0002 & 0.0006 & 0.048 & 0.048 & 0.0003 & 0.0006 & 0.056 & 0.058 \\ \cline{3-12}
 &  & \multirow{3}{*}{$1/5$} & $\hat{\tau}^{MI}$ & 0.0166 & 0.0153 & 0.446 & 0.426 & 0.0160 & 0.0128 & 0.444 & 0.422 \\
 &  &  & $\hat{\tau}^{NP}$ & 0.0249 & 0.0090 & 0.666 & 0.674 & -0.0008 & 0.0080 & 0.590 & 0.600 \\
 &  &  & $\hat{\tau}^{RP}$ & -0.1370 & 0.0195 & 0.762 & 0.764 & -0.1488 & 0.0229 & 0.564 & 0.564 \\ \cline{3-12}
 &  & \multirow{3}{*}{$1/3$} & $\hat{\tau}^{MI}$ & 0.0250 & 0.0142 & 0.848 & 0.836 & 0.0237 & 0.0111 & 0.874 & 0.864 \\
 &  &  & $\hat{\tau}^{NP}$ & 0.0324 & 0.0089 & 0.968 & 0.970 & -0.0083 & 0.0067 & 0.950 & 0.946 \\
 &  &  & $\hat{\tau}^{RP}$ & -0.2321 & 0.0548 & 0.986 & 0.984 & -0.2520 & 0.0642 & 0.948 & 0.946 \\ \hline

\end{tabular}
}
\end{table}

\begin{table}[htb!]
\centering
\caption{Biases, mean squared error (MSE), 
empirical power evaluated by asymptotic null distribution (EP-A), and
empirical power evaluated by permutation (EP-P) for bivariate interval-censored data under the given scenario {\bf (C2)}.} \medskip
\scalebox{0.95}{
\begin{tabular}{cccccccccccc} \hline
\multirow{2}{*}{$c_R$} &\multirow{2}{*}{$n$}  & \multirow{2}{*}{$\tau$} &\multirow{2}{*}{Est.} &
\multicolumn{4}{c}{Clayton} & \multicolumn{4}{c}{Frank} \\ \cline{5-12}
&  &  &  & Bias & MSE & EP-A & EP-P & Bias & MSE & EP-A & EP-B \\ \hline
\multirow{18}{*}{$6$} & \multirow{9}{*}{$50$} & \multirow{3}{*}{$0$} & $\hat{\tau}^{MI}$ & 0.0014 & 0.0054 & 0.058 & 0.068 & 0.0002 & 0.0053 & 0.066 & 0.074 \\
 &  &  & $\hat{\tau}^{NP}$ & 0.0033 & 0.0151 & 0.064 & 0.064 & 0.0015 & 0.0158 & 0.046 & 0.046 \\
 &  &  & $\hat{\tau}^{RP}$ & 0.0023 & 0.0077 & 0.064 & 0.064 & 0.0011 & 0.0080 & 0.054 & 0.056 \\ \cline{3-12}
 &  & \multirow{3}{*}{$1/5$} & $\hat{\tau}^{MI}$ & -0.0920 & 0.0137 & 0.276 & 0.320 & -0.0873 & 0.0131 & 0.332 & 0.368 \\
 &  &  & $\hat{\tau}^{NP}$ & 0.0289 & 0.0149 & 0.454 & 0.446 & 0.0316 & 0.0151 & 0.472 & 0.474 \\
 &  &  & $\hat{\tau}^{RP}$ & -0.0380 & 0.0084 & 0.456 & 0.452 & -0.0354 & 0.0084 & 0.476 & 0.486 \\ \cline{3-12}
 &  & \multirow{3}{*}{$1/3$} & $\hat{\tau}^{MI}$ & -0.1504 & 0.0290 & 0.700 & 0.718 & -0.1398 & 0.0248 & 0.762 & 0.782 \\
 &  &  & $\hat{\tau}^{NP}$ & 0.0390 & 0.0146 & 0.864 & 0.868 & 0.0484 & 0.0140 & 0.896 & 0.902 \\
 &  &  & $\hat{\tau}^{RP}$ & -0.0687 & 0.0113 & 0.864 & 0.866 & -0.0622 & 0.0097 & 0.896 & 0.892 \\ \cline{2-12}
 & \multirow{9}{*}{$100$} & \multirow{3}{*}{$0$} & $\hat{\tau}^{MI}$ & 0.0025 & 0.0025 & 0.050 & 0.058 & 0.0012 & 0.0026 & 0.052 & 0.060 \\
 &  &  & $\hat{\tau}^{NP}$ & 0.0003 & 0.0071 & 0.044 & 0.042 & -0.0023 & 0.0073 & 0.048 & 0.046 \\
 &  &  & $\hat{\tau}^{RP}$ & 0.0002 & 0.0036 & 0.046 & 0.046 & -0.0016 & 0.0037 & 0.050 & 0.050 \\ \cline{3-12}
 &  & \multirow{3}{*}{$1/5$} & $\hat{\tau}^{MI}$ & -0.0950 & 0.0113 & 0.578 & 0.618 & -0.0891 & 0.0104 & 0.584 & 0.632 \\
 &  &  & $\hat{\tau}^{NP}$ & 0.0230 & 0.0074 & 0.752 & 0.752 & 0.0326 & 0.0081 & 0.790 & 0.794 \\
 &  &  & $\hat{\tau}^{RP}$ & -0.0418 & 0.0052 & 0.754 & 0.758 & -0.0350 & 0.0047 & 0.788 & 0.780 \\ \cline{3-12}
 &  & \multirow{3}{*}{$1/3$} & $\hat{\tau}^{MI}$ & -0.1549 & 0.0265 & 0.946 & 0.946 & -0.1463 & 0.0238 & 0.958 & 0.964 \\
 &  &  & $\hat{\tau}^{NP}$ & 0.0418 & 0.0082 & 0.996 & 0.996 & 0.0554 & 0.0089 & 0.994 & 0.994 \\
 &  &  & $\hat{\tau}^{RP}$ & -0.0668 & 0.0077 & 0.996 & 0.996 & -0.0574 & 0.0062 & 0.994 & 0.994 \\ \hline
\multirow{18}{*}{$12$} & \multirow{9}{*}{$50$} & \multirow{3}{*}{$0$} & $\hat{\tau}^{MI}$ & -0.0056 & 0.0094 & 0.048 & 0.056 & 0.0037 & 0.0076 & 0.040 & 0.038 \\
 &  &  & $\hat{\tau}^{NP}$ & -0.0045 & 0.0144 & 0.058 & 0.062 & 0.0045 & 0.0120 & 0.042 & 0.042 \\
 &  &  & $\hat{\tau}^{RP}$ & -0.0036 & 0.0094 & 0.054 & 0.058 & 0.0037 & 0.0079 & 0.044 & 0.038 \\ \cline{3-12}
 &  & \multirow{3}{*}{$1/5$} & $\hat{\tau}^{MI}$ & -0.0450 & 0.0100 & 0.364 & 0.336 & -0.0226 & 0.0085 & 0.486 & 0.486 \\
 &  &  & $\hat{\tau}^{NP}$ & 0.0055 & 0.0112 & 0.424 & 0.418 & 0.0265 & 0.0117 & 0.522 & 0.524 \\
 &  &  & $\hat{\tau}^{RP}$ & -0.0342 & 0.0084 & 0.424 & 0.422 & -0.0166 & 0.0074 & 0.524 & 0.522 \\ \cline{3-12}
 &  & \multirow{3}{*}{$1/3$} & $\hat{\tau}^{MI}$ & -0.0577 & 0.0107 & 0.858 & 0.858 & -0.0382 & 0.0081 & 0.902 & 0.908 \\
 &  &  & $\hat{\tau}^{NP}$ & 0.0313 & 0.0117 & 0.902 & 0.906 & 0.0422 & 0.0108 & 0.932 & 0.938 \\
 &  &  & $\hat{\tau}^{RP}$ & -0.0394 & 0.0084 & 0.902 & 0.906 & -0.0295 & 0.0067 & 0.932 & 0.938 \\ \cline{2-12}
 & \multirow{9}{*}{$100$} & \multirow{3}{*}{$0$} & $\hat{\tau}^{MI}$ & -0.0042 & 0.0037 & 0.034 & 0.034 & 0.0051 & 0.0042 & 0.064 & 0.062 \\
 &  &  & $\hat{\tau}^{NP}$ & -0.0047 & 0.0061 & 0.036 & 0.036 & 0.0082 & 0.0069 & 0.062 & 0.066 \\
 &  &  & $\hat{\tau}^{RP}$ & -0.0038 & 0.0040 & 0.036 & 0.038 & 0.0066 & 0.0045 & 0.064 & 0.064 \\ \cline{3-12}
 &  & \multirow{3}{*}{$1/5$} & $\hat{\tau}^{MI}$ & -0.0552 & 0.0069 & 0.654 & 0.648 & -0.0315 & 0.0046 & 0.768 & 0.756 \\
 &  &  & $\hat{\tau}^{NP}$ & 0.0107 & 0.0061 & 0.770 & 0.762 & 0.0280 & 0.0065 & 0.820 & 0.820 \\
 &  &  & $\hat{\tau}^{RP}$ & -0.0297 & 0.0048 & 0.766 & 0.770 & -0.0157 & 0.0040 & 0.816 & 0.810 \\ \cline{3-12}
 &  & \multirow{3}{*}{$1/3$} & $\hat{\tau}^{MI}$ & -0.0738 & 0.0090 & 0.986 & 0.984 & -0.0479 & 0.0059 & 0.996 & 0.996 \\
 &  &  & $\hat{\tau}^{NP}$ & 0.0227 & 0.0055 & 0.996 & 0.996 & 0.0482 & 0.0075 & 1.000 & 1.000 \\
 &  &  & $\hat{\tau}^{RP}$ & -0.0460 & 0.0053 & 0.996 & 0.996 & -0.0250 & 0.0039 & 1.000 & 1.000 \\ \hline
\end{tabular}
}
\end{table}

\begin{table}[htb!]
\centering
\caption{Summary of the independence test for the leukemia remission,
animal tumorigenicity, and ACTG 181 datasets}\label{tab:app}
\medskip
\scalebox{0.95}
{
\begin{tabular}{| c | ccc | ccc | ccc|}\hline
\multirow{2}{*}{Test stat.} & \multicolumn{3}{|c|}{Leukemia remission}  & \multicolumn{3}{|c|}{Animal tumorigenicity} & \multicolumn{3}{|c|}{ACTG 181} \\\cline{2-10}
 & Estimate & EP-A & EP-P & Estimate & EP-A & EP-P & Estimate & EP-A & EP-P \\ \hline
$\hat{\tau}^{O}$ & -0.0667 & 0.5680 & 0.5748 & - & - & - & - & - & - \\
$\hat{\tau}^{MI}$ & - & - & - & 0.0955 & 0.2851 & 0.2046 & -0.0654 & 0.2260 & 0.1706 \\
$\hat{\tau}^{NP}$ & -0.0652 & 0.7163 & 0.7200 & 0.1689 & 0.1853 & 0.1915 & -0.1035 & 0.0957 & 0.0960 \\
$\hat{\tau}^{RP}$ & -0.0540 & 0.6787 & 0.6998 & 0.0093 & 0.2519 & 0.2641 & -0.0270 & 0.0962 & 0.0954 \\
\hline
\end{tabular}
}
\end{table}

%\section*{Acknowledgement} 
%J. Lim's research is supported by National Research Foundation of Korea (No. NRF-2017R1A2B2012264)

\section*{Appendix A: Proof of Proposition 1 and Proposition 2} 

The proof of Proposition 2 is almost same with that of Proposition 1. In below, we only give the proof of Proposition 1. 

For uncensored observations, the minimum possible rank of $x_i=x_i^*$ for $i\in I_U$
is equal to the rank of $x_i$ among $(x_j)_{j \in I_U}$
since the unobserved random variable $X_j^*$ for $j \in I_R$ can be 
 greater than $x_i$ even if $x_j = x_j^{c} \le x_i$ under the condition $X_j^* > x_j$
for $j \in I_R$. Here, the rank of $x_i$ is defined as $\sum_{j\in I_U} I(x_j<x_i) + 1$,
which is equivalent to the minimum rank of $x_i$ when there are ties.
Let $t = \sum_{j\in I_U} I(x_j=x_i)$ be the number of observations equal to $x_i$ in $(x_j)_{j\in I_U}$.
From the minimum possible rank of $x_i$, the rank of $x_i$ can be increased by
$t-1$ plus the number of right-censored observations less than $x_i$.
The maximum possible rank of $x_i^*$ for $j\in I_U$ can be represented by  
$\sum_{j\in I_U} I(x_j<x_i) + t + \sum_{j\in I_R} I(x_j < x_i^*)$.
Thus, the set of possible ranks of $x_i^*$ is $\big\{
R : \sum_{j\in U} I(x_j<x_i) + 1 \le R \le 
\sum_{j\in I_U} I(x_j\le x_i) + \sum_{j\in I_R} I(x_j < x_i) \big\}$.
Similarly, for right-censored observations,
the minimum possible ranks of $x_i=x_i^{c}$ for $i\in I_R$ is
$\sum_{j\in U} I(x_j\le x_i) + 1$, which is increased by $\sum_{j\in I_U} I(x_j=x_i)$
from the form of the minimum possible rank of the uncensored observation due to the condition $X_i^* > x_i$.
The maximum possible rank of $x_i$ is equal to $n$
since the ranks of right-censored observations are exchangeable
in the sense that either $X_{j}^* < X_{l}^*$ or 
$X_{j}^*> X_{l}^*$ is possible for $j,l \in I_R$, $X_{j}^* \in (x_{j}^{c}, \infty)$, and $X_{l}^* \in (x_{l}^{c}, \infty)$.
Thus, the set of possible ranks of $x_{i} = x_{i}^{c}$
is $\big\{R : \sum_{j\in U} I(x_j\le x_i) + 1 \le R \le n  \big\}$.

\section*{Appendix B: Results of the numerical study for parallel tempering}

\begin{figure}
\centering
\includegraphics[width=\linewidth]{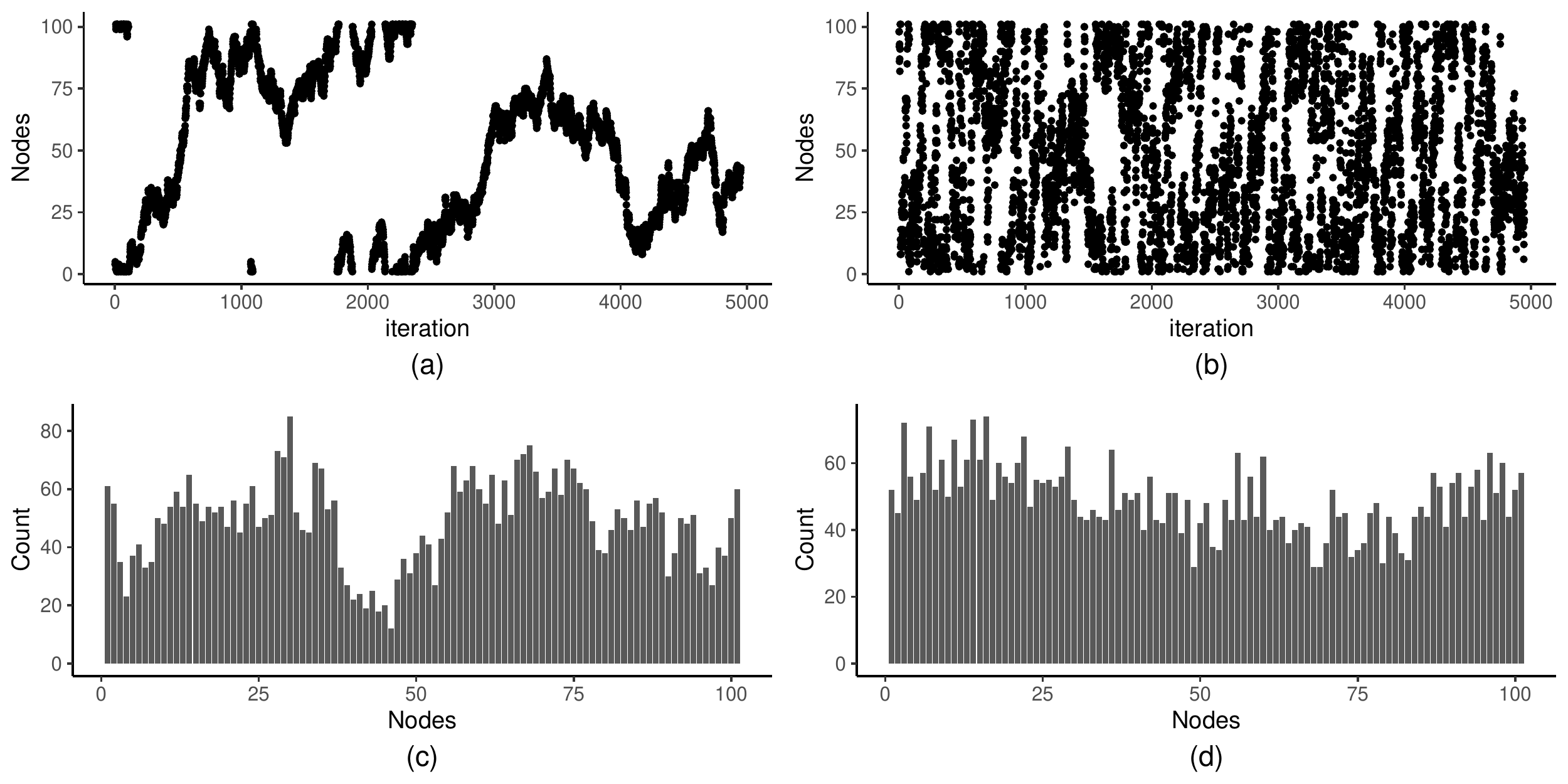}
\caption{Scatter plot (top) and histogram (bottom) of the target chain on the circulant graph $C_{101}^{1}$. Plot (a) and (c) represent the behavior of the target chain for the case in which two chains are independent whereas plot (b) and (d) show the behavior of the target chain for the case in which two chains are Metropolis-coupled.}\label{fig:ar1_sh}
\end{figure}

\begin{figure}
\centering
\includegraphics[width=\linewidth]{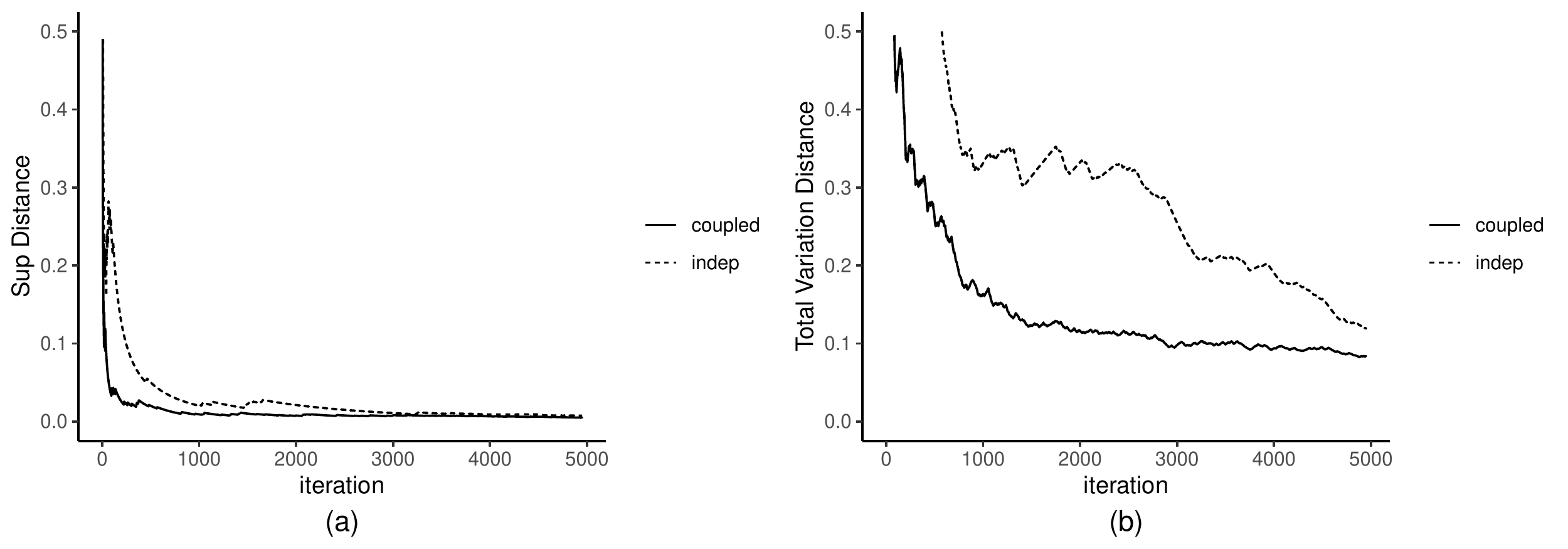}
\caption{Plots of distance changes of the target chains on the circulant graph $C_{101}^{1}$ : the left one represents the convergence of $L_{\infty}$ distance and the right one shows the convergence of total variation distance. The solid lines represent the distance change of the coupled target chain, and the dotted lines stand for the distance change of the independent target chain.}\label{fig:ar1_d}
\end{figure}

\begin{figure}
\centering
\includegraphics[width=\linewidth]{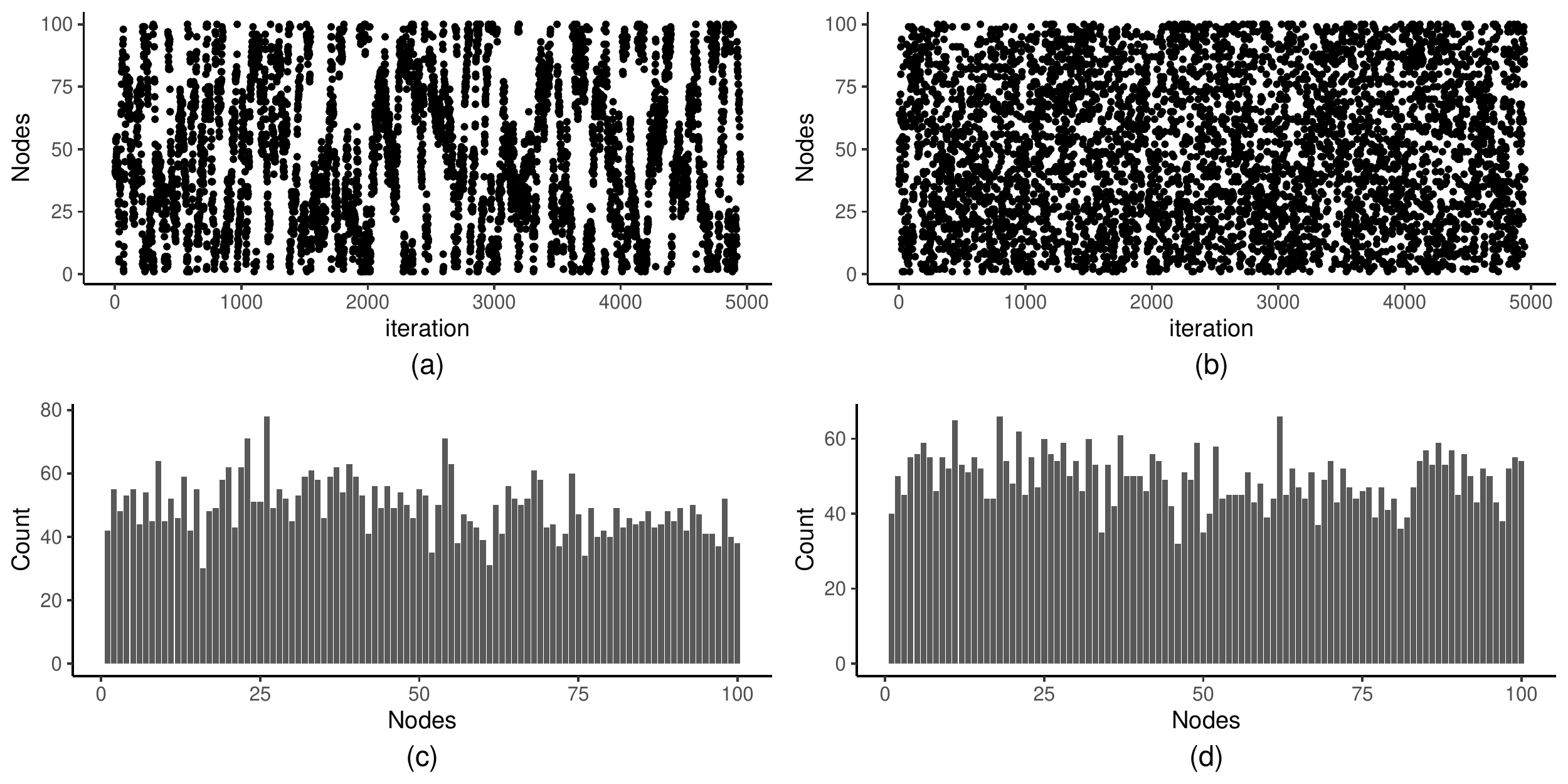}
\caption{Scatter plot (top) and histogram (bottom) of the target chain on the circulant graph $C_{100}^{1,2,3}$. Plot (a) and (c) represent the behavior of the target chain for the case in which two chains are independent whereas plot (b) and (d) show the behavior of the target chain for the case in which two chains are Metropolis-coupled.}\label{fig:ar3_sh}
\end{figure}

\begin{figure}
\centering
\includegraphics[width=\linewidth]{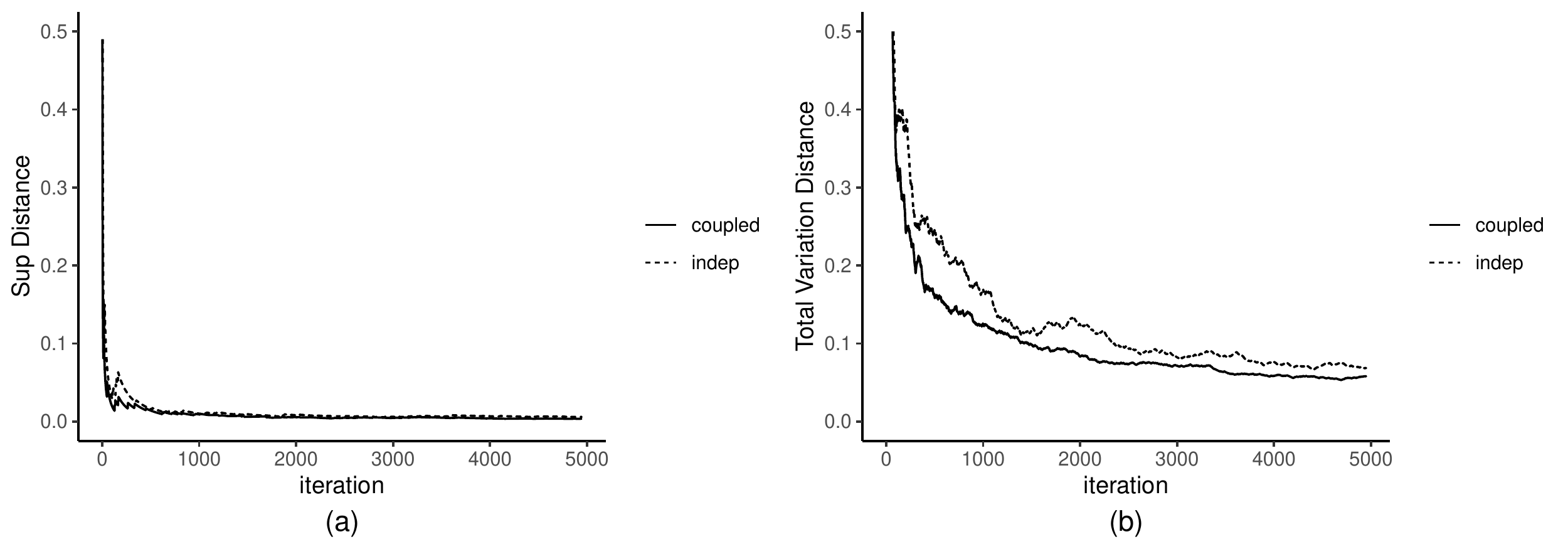}
\caption{Plots of distance changes of the target chains on the circulant graph $C_{100}^{1,2,3}$ : the left one represents the convergence of $L_{\infty}$ distance and the right one shows the convergence of total variation distance. The solid lines represent the distance change of the coupled target chain, and the dotted lines stand for the distance change of the independent target chain.}\label{fig:ar3_d}
\end{figure}

\begin{figure}
\centering
\includegraphics[width=\linewidth]{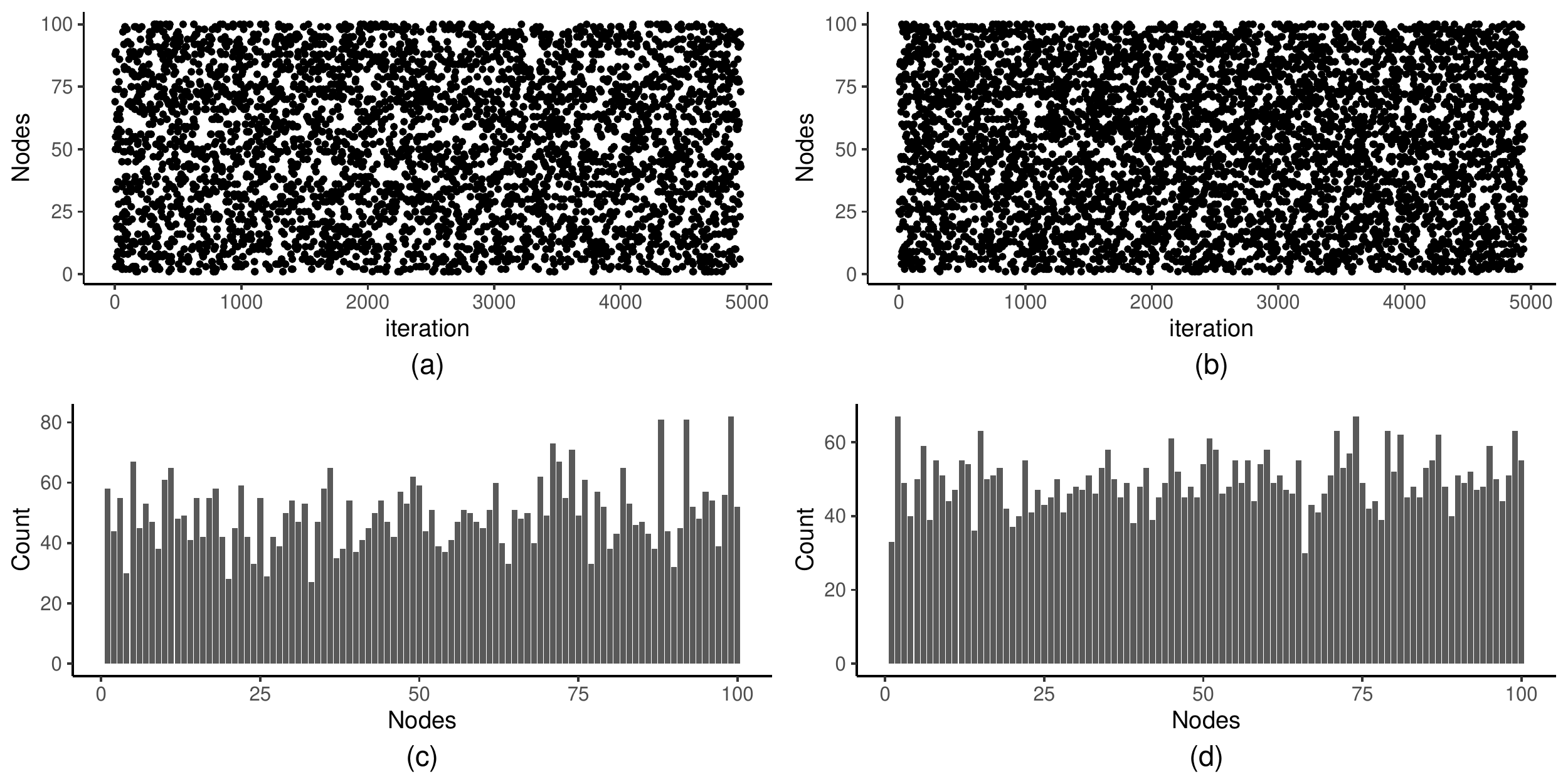}
\caption{Scatter plot (top) and histogram (bottom) of the target chain on the random 3-regular graph. Plot (a) and (c) represent the behavior of the target chain for the case in which two chains are independent whereas plot (b) and (d) show the behavior of the target chain for the case in which two chains are Metropolis-coupled.}\label{fig:reg_sh}
\end{figure}

\begin{figure}
\centering
\includegraphics[width=\linewidth]{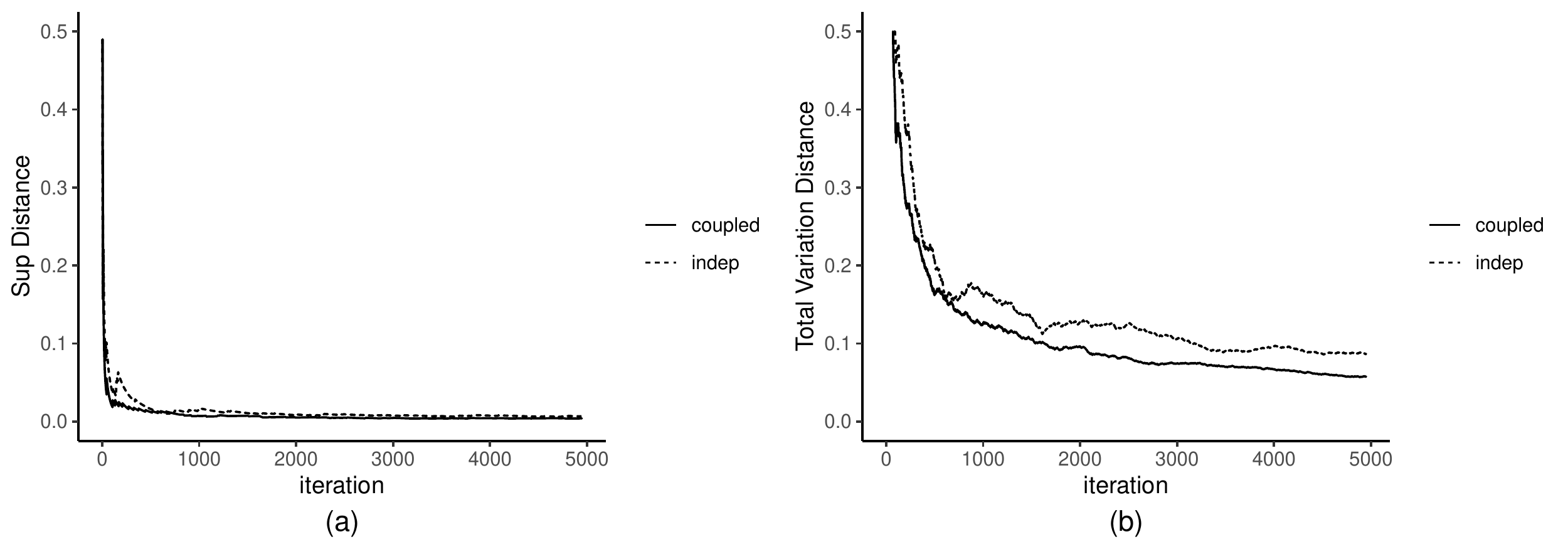}
\caption{Plots of distance changes of the target chains on the random 3-regular graph : the left one represents the convergence of $L_{\infty}$ distance and the right one shows the convergence of total variation distance. The solid lines represent the distance change of the coupled target chain, and the dotted lines stand for the distance change of the independent target chain.}\label{fig:reg_d}
\end{figure}

\begin{figure}
\centering
\includegraphics[width=\linewidth]{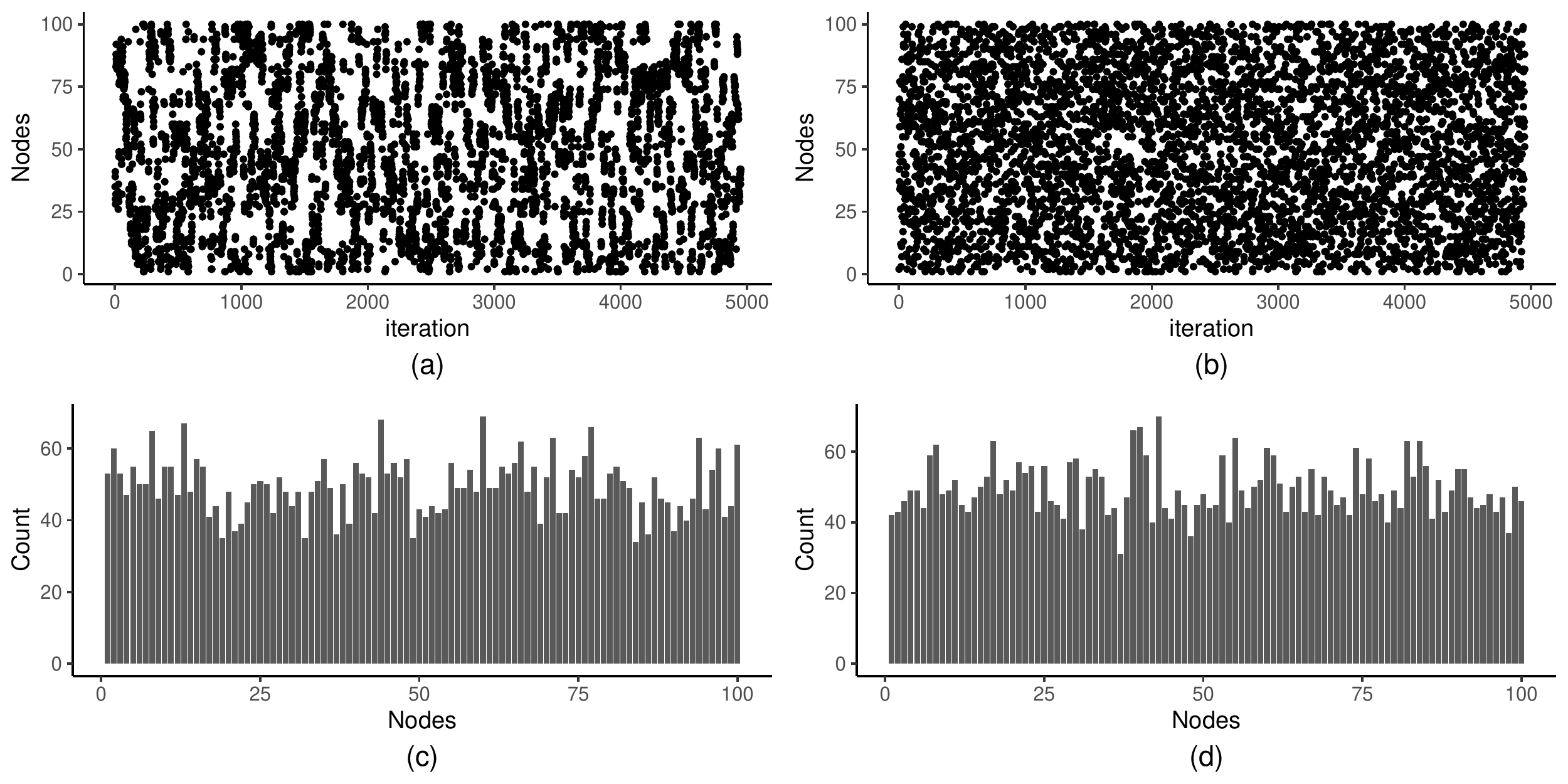}
\caption{Scatter plot (top) and histogram (bottom) of the target chain on the Watts–Strogatz model. Plot (a) and (c) represent the behavior of the target chain for the case in which two chains are independent whereas plot (b) and (d) show the behavior of the target chain for the case in which two chains are Metropolis-coupled.}\label{fig:sw_sh}
\end{figure}

\begin{figure}
\centering
\includegraphics[width=\linewidth]{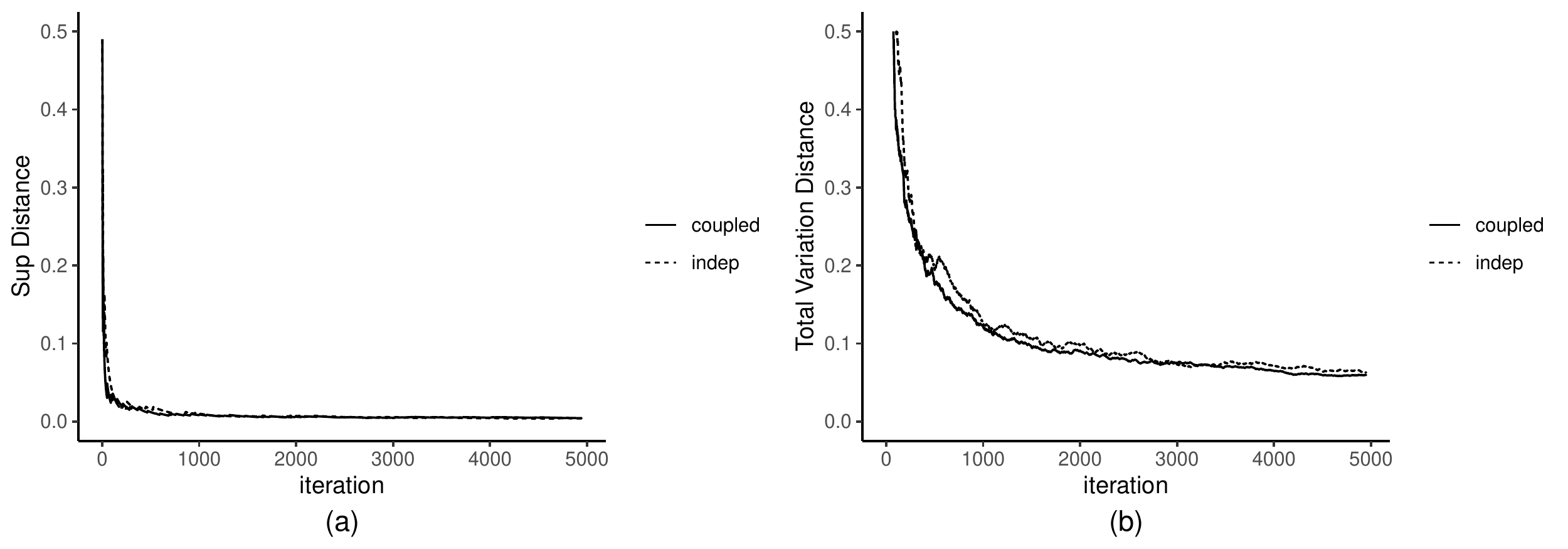}
\caption{Plots of distance changes of the target chains on the Watts-Strogatz model : the left one represents the convergence of $L_{\infty}$ distance and the right one shows the convergence of total variation distance. The solid lines represent the distance change of the coupled target chain, and the dotted lines stand for the distance change of the independent target chain.}\label{fig:sw_d}
\end{figure}

Figures \ref{fig:ar1_sh}, \ref{fig:ar3_sh}, \ref{fig:reg_sh} and \ref{fig:sw_sh} show the difference between the behaviors of target chains. We can see some patterns in the scatter plots of the independent target chains while the scatter plots of the coupled target chains have relatively less patterns. The histograms also imply that the distribution of coupled target chains is closer to the uniform distribution than that of independent target chains. Figures \ref{fig:ar1_d}, \ref{fig:ar3_d}, \ref{fig:reg_d} and \ref{fig:sw_d} show that the convergence of the distribution of target chain to the limiting distribution is accelerated by the parallel tempering. Therefore, we can conclude that the Metropolis-coupling makes target chain travel more evenly over the nodes.

\end{document}